\documentclass[11pt,journal,draftcls,letterpaper,onecolumn]{IEEEtran}
\usepackage[tbtags]{amsmath}
\usepackage{amssymb}
\usepackage{bm}
\usepackage[font=bf]{subfig}
\usepackage{xspace}
\usepackage{mathrsfs}
\usepackage{cite}
\usepackage{geometry}
\usepackage[dvips]{graphicx}
\usepackage{xspace}
\usepackage{srcltx}
\usepackage{color}
\newcommand{\sign}{\textrm{sign}}
\usepackage{tabularx}
\usepackage{algorithm}
\usepackage{algorithmic}
\usepackage{enumerate}

\title{Convex Feasibility Methods for Compressed Sensing }

\author{Avishy Carmi and
        Pini Gurfil,~\IEEEmembership{Member,~IEEE}
\thanks{Manuscript received ; revised .}
\thanks{Copyright (c) 2010 IEEE. Personal use of this material is permitted. However, permission to use this material for any other purposes must be obtained from the IEEE by sending a request to pubs-permissions@ieee.org.}
\thanks{A. Carmi is with the Asher Space Research Institute, Technion -- Israel Institute of Technology, Haifa 32000, Israel. }
\thanks{P. Gurfil is with the Faculty of Aerospace Engineering, Technion -- Israel Institute of Technology, Haifa 32000, Israel.}
}

\date{}

\renewcommand{\include}{\input}

\begin{document}

\maketitle

\begin{abstract}
We present a computationally-efficient method for recovering sparse signals from a series of noisy observations, known as the problem of compressed
sensing (CS). CS theory requires solving a convex constrained minimization problem. We propose to transform this optimization problem into a convex
feasibility problem (CFP), and solve it using subgradient projection methods, which are iterative, fast, robust and convergent schemes for solving CFPs.
As opposed to some of the recently-introduced CS algorithms, such as Bayesian CS and gradient projections for sparse reconstruction, which
become prohibitively inefficient as the problem dimension and sparseness degree increase, the newly-proposed methods exhibit a marked
robustness with respect to these factors. This renders the subgradient projection methods highly viable for large-scale compressible
scenarios.
\end{abstract}


\section {Introduction}

Recent studies have shown that sparse signals can be recovered accurately using less observations than
predicted by the Nyquist/Shannon sampling principle; the resulting theory is known as compressed sensing (CS)~\cite{cs:06, candes:06}.
The essence of the new theory builds upon a new data acquisition formalism, in which compression plays a fundamental role.
Sparse -- and more generally -- compressible signals arise naturally in many fields of science and engineering, such as the reconstruction of
images from under-sampled Fourier data, biomedical imaging and astronomy~\cite{MRI:07, MRI:08}.
Other applications include model-reduction methods to enforce sparseness for preventing over-fitting and for reducing computational
complexity and storage capacities. The reader is referred to Refs.~\cite{cs:06} and \cite{candes:06} for
an extensive overview of CS theory.

The recovery of sparse signals consists of solving an NP-hard minimization problem~\cite{cs:06, nonconvex:07}.
State-of-the-art methods for addressing this optimization problem commonly utilize convex relaxations,
non-convex local optimization and greedy search mechanisms.
Convex relaxations are used in various methods such as least absolute shrinkage and selection operator (LASSO) \cite{tibshirani:96}, least
angle regression (LARS) \cite{lars:04},
the Dantzig selector (DS) \cite{dantzig:06}, basis pursuit (BP) and basis pursuit de-noising \cite{bp:98}. The two
former schemes, the LASSO and LARS, are essentially homotopy methods that exploit pivoting operations on sub-matrices
from the complete sensing matrix (sub-sensing matrices) for yielding a solution path to the convex optimization problem. These methods
turn out to be extremely efficient whenever the sparseness level is relatively high owing to the fact that only
a few sub-sensing matrices need to be provided, corresponding to the instantaneous support of the underlying
reconstructed signal. The latter methods, the DS and the BP variants, recast a linear program and employ
either simplex or interior-point techniques for obtaining an optimal solution. Similarly to the homotopy-based approaches, these methods may become computationally-intensive, since the number of elements in the support of the underlying signal
increases.

Non-convex optimization approaches include Bayesian methodologies such as the relevance vector machine (RVM), otherwise known as
sparse Bayesian learning~\cite{rvm:01} as well as stochastic search algorithms, which are mainly based on Markov chain Monte
Carlo (MCMC) techniques~\cite{mc1,mc2,mc3,mc4}. In virtue of their Bayesian mechanism and in contrast to other
optimization approaches, these methods provide a complete statistical solution to the problem by means of a probability
density function. Nevertheless, the intensive computational requirements of these  methods render their applicability
questionable in high-dimensional problems.

Recently, the Bayesian framework was utilized to create efficient CS schemes~\cite{bcs,cskf}. The
Bayesian CS algorithm in \cite{bcs} exploits both a sparseness-promoting hierarchical prior and an RVM mechanism for deriving
point estimates and statistical error bounds. This method was shown to outperform some of the commonly-used greedy schemes
both in accuracy and speed. The work in \cite{cskf} derived a pseudo-measurement-based Kalman filtering algorithm (CSKF)
for recovering sparse signals from noisy observations in a sequential manner. This approach extends CS to accommodate stochastic
linear filtering problems and, similarly to the aforementioned Bayesian methods, yields a complete statistical solution to the problem
(a Gaussian distribution).

Notable greedy search algorithms are the matching pursuit (MP)~\cite{mp}, the orthogonal MP~\cite{omp},
and the orthogonal least squares (OLS)~\cite{ols}. Both MP and OMP minimize the reconstruction error
by iteratively choosing elements from the dictionary matrix (sensing matrix). The OMP
involves an additional orthogonalization stage and is known to outperform the conventional MP.
The OLS works in a similar fashion, employing an orthogonal transformation of the original sensing matrix.
The greedy algorithms are known to posses an advantage in terms of computational cost over other optimization schemes when
the basis projections are sufficiently incoherent and the number of elements in the support is relatively small.
This, in turn, implies that their performance may deteriorate dramatically in common realistic settings wherein the
underlying signals are compressible rather than sparse.

The scalability of a CS technique essentially refers to its viability for high dimensional settings. This issue is given a prime consideration in \cite{gp}, where a new gradient projection (GP)-based method is proposed for solving
possibly large-scale CS problems. As pointed out in \cite{gp}, large-scale methods require only matrix-vector
products involving the sensing matrix. As demonstrated in \cite{gp}, the GP algorithm
is efficient in high-dimensional settings compared to the aforementioned methods. However,
similarly to the homotopy methods, the practical implementation of the GP algorithm requires a delicate tuning of the $l_1$-norm bounding parameter, which greatly affects its convergence and optimality.

\subsection{Restricted Isometry Property and Convex Relaxations}

The theory of compressed sensing has drawn much attention to the convex relaxation methods.
It has been shown that the convex $l_1$ relaxation yields an exact solution
to the recovery problem provided two conditions are met: 1) the signal is sufficiently sparse, and
2) the sensing matrix obeys the so-called \emph{restricted isometry property} (RIP) at a certain level. Another complementary
result ensures high accuracy when dealing with noisy observations. Further elaboration of this result facilitated
its probabilistic version, which is concluded by the known statement of recovery `with overwhelming probability'~\cite{cs:06}.
To put it informally, it is highly probable for the convex $l_1$ relaxation to yield an exact solution
provided the involved quantities, the sparseness degree $s$, and the sensing matrix dimensions $m \times n$
maintain a relation of the type $s = \mathcal{O}(m / \log (n/m))$.


\subsection{Convex Feasibility Problems, Subgradient Projections, and Compressed Sensing}

The issue of computational efficiency is of prime importance for large-scale CS problems. Ideally, a computational scheme for recovering
sparse and/or compressible signals should be characterized by (i) computational efficiency; (ii) minimum mean-square error of the estimated signals;
and (iii) scalability. Additional desirable features would be simple and transplant implementation (no ``black boxes"); robustness with respect to
some sparsensess/compressibility measure; and ubiquitous applicability to real-world problems.

A potentially promising approach for satisfying the aforementioned goals is to transform the CS problem into a convex feasibility problem (CFP).
CFPs are aimed at finding a point in the intersection set of nonlinear convex sets. The CFP is a fundamental problem in numerous fields,
see, e.~g., \cite{c93,c96} and references therein. It has been used to model real-world problems in image reconstruction from
projections \cite{H80}, in radiation therapy treatment planning \cite{cap88}, and in crystallography \cite{msl99}.

Most often, CFPs are solved by performing projections onto the individual sets. This is carried out in various ways, by using different
projection methods, resulting in a myriad of algorithms that usually exhibit good convergence while consuming minimal computational resources.
Some of the projection algorithms may even provide predictable performance when the solution set of the CFP is empty \cite{butnariu}.  We refer
the reader to Ref.~\cite{yamada} for a discussion on the connection between projection methods and variational inequalities;
to Refs.~\cite{crombez03, crombez06} for applications in signal processing; and to \cite[Chapter 5]{CZ97} for a comprehensive review.

However, projections onto convex sets may be prohibitively difficult when the constraints are nonlinear. This is so because computing orthogonal
projections onto arbitrary convex sets requires a separate, inner-loop minimization for finding the minimum distance between a given point
and an arbitrary curve, a process that usually involves a considerable computational effort.

An elegant alternative is to use \emph{subgradient projections}. Subgradient projections only require the instantaneous subgradient in order to
perform the next iteration. The projection is performed onto an intermediate point, and not directly onto the convex set. Some of the
subgradient projection methods, such as cyclic subgradient projections \cite{lc82} have a convergence proof when the intersection set of all the
convex constraints is non-empty.

In this paper, we use both cyclic and simultaneous subgradient projections (CSP and SSP) for efficiently solving the CS problem.
The proposed algorithms are characterized by
(i) computational efficiency; (ii) accuracy; (iii) robustness to varying compressibility and sparseness levels; and (iv) straightforward,
transparent implementation. By using extensive numerical evaluations, we illustrate the superiority of the proposed scheme --
in terms of computational efficiency and accuracy -- compared to commonly-used methods for solving the CS problem.


\section {Sparse Signal Recovery and Compressed Sensing}
Consider a signal $\chi \in \mathbb{R}^n$ that is sparse in some domain, i.e., it can be represented
using a relatively small number of projections in some known, possibly orthonormal, basis, say $\psi \in \mathbb{R}^{n \times n}$.
Thus, we may write
\begin{equation}
\label{cs:signal}
\chi = \psi x = \sum^n_{i=1} x_i \psi_i = \sum_{x_j \in \mathrm{supp}(x)} x_j \psi_j, \qquad \| x \|_0 < n
\end{equation}
where $\mathrm{supp}(x)$ and $\| x \|_0$ (the $l_0$ norm) denote the support of $x$ and its dimension (i.~e., the number of non-zero
elements of $x$), respectively. The problem of compressed sensing considers the recovery of $x$ (and therefore of $\chi$) from a limited number,
$m < n$, of incoherent and possibly noisy measurements (or, in other words, sensing a compressible signal from a limited
number of incoherent measurement)~\cite{cs:06}. The measurements themselves obey a linear relation of the form
\begin{equation}
\label{cs:obsv}
y = H' \chi = H x
\end{equation}
where $H \in \mathbb{R}^{m \times n}$ and $H=H'\psi$. In many practical applications, the observation vector $y$ may be either inaccurate
or contaminated by noise. In this case, which will be referred to as the \emph{stochastic CS problem}, an additional noise term
is added to the right-hand side of \eqref{cs:obsv}.

In general, under certain limitations on the sparseness degree of $x$, $s=\| x \|_0$,
an exact solution to the above recovery problem can be obtained by essentially solving a subset selection problem of the form
\begin{equation}
\label{eq:p1}
\min_{\hat x} \| \hat x \|_0 \text{ s.t. } \| y-H \hat x \|^2_2 \le \epsilon
\end{equation}
for a sufficiently small $\epsilon$. However, the problem \eqref{eq:p1} is known to be NP-hard, which implies
that in practice an optimizer $\hat x$ cannot be computed efficiently.

In the late 90's, the $l_1$ norm was suggested as a sparseness-promoting term in the seminal
work introducing the acclaimed LASSO operator~\cite{tibshirani:96} and the basis pursuit
\cite{bp:98}. Recasting the sparse recovery problem using the $l_1$ norm yields a convex relaxation of the original NP-hard
problem, which can be efficiently solved via a myriad of well-established optimization techniques.
Commonly, there are two equivalent convex formulations that follow from \eqref{eq:p1}: The quadratically-constrained linear program, which takes the form
\begin{equation}
\label{eq:c1}
\min_{\hat x} \| \hat x \|_1 \text{ s.t. } \| y-H \hat x \|^2_2 \le \epsilon
\end{equation}
and the quadratic program
\begin{equation}
\label{eq:c2}
\min_{\hat x} \| y-H \hat x \|^2_2 \text{ s.t. } \| \hat x \|_1 \le \epsilon'
\end{equation}
It can be shown that for proper values of the tuning parameters $\epsilon$ and $\epsilon'$ the solution of both these
problems coincide.

Recently,\cite{cs:06, candes:06} have shown that an accurate solution of \eqref{eq:p1} can
almost always be obtained by solving the convex relaxation \eqref{eq:c1} assuming that the sensing matrix
$H$ obeys the RIP mentioned before. The RIP roughly implies that the columns of a
given matrix nearly behave like an orthonormal basis. This desired property is possessed by several random constructions,
which guarantee the uniqueness of the sparse solution. In particular, an exact recovery is highly probable when
using such matrices provided that a relation of the type
\begin{equation}
s=\mathcal{O}(m/\log(n/m))
\end{equation}
holds. For an extensive overview of several RIP constructions and their role in CS, the reader is referred to \cite{cs:06,candes:06}.


In practice, the projected signal $x$ may be \emph{nearly} sparse, in the sense of having many relatively small elements, which are not identically
zero. Such representations, frequently encountered in real-world applications, are termed \emph{compressible}. Most of the results in the CS literature naturally extend to the compressible case assuming some behavior of the
small nonzero elements. Such a behavior is suggested in \cite{candes:06}, where the compressible element sequence is assumed to decay
according to the power law
\begin{equation}
|x_i| \le \kappa i^{-1/r}, \qquad |x_i| \ge |x_{i+1}|
\end{equation}
where $\kappa > 0$ and $r > 0$ are the radius of a weak $l_r$ ball to which $x$ is confined, and a decay factor, respectively.
In this case, an equivalent measure of the signal sparseness degree, $s$, can be obtained as
\begin{equation}
\label{eq:sd}
\hat s= n - \mathrm{card}\{i \mid |x_i| \le \varepsilon \}
\end{equation}
for some sufficiently small $\varepsilon > 0$, where $\mathrm{card}\{\cdot\}$ denotes the cardinality of a set.


\section{Subgradient Projections for Compressed Sensing}

In this section, we outline the convex feasibility problem and two algorithms for a solution thereof: The cyclic subgradient projections (CSP) and the simultaneous subgradient projections (SSP). We will explain how these algorithms and their variants, collectively referred to as \emph{convex feasibility methods}, are implemented for efficiently solving the convex CS problem described by Eqs.~\eqref{eq:c1} or \eqref{eq:c2}.

\subsection{The Convex Feasibility Problem}

Given $p$ closed convex subsets $Q_{1},Q_{2},\cdots,Q_{p}\subseteq \mathbb R^{n}$ of
the $n$-dimensional Euclidean space, expressed as%
\begin{equation}
Q_{i}=\left\{  z\in \mathbb R^{n}\mid f_{i}(z)\leq0\right\}  , \label{eq:set}%
\end{equation}
where $f_{i}:\mathbb R^{n}\rightarrow \mathbb R$ is a convex function, the \textit{convex
feasibility problem} (CFP) is%
\begin{equation}
\text{find a point }z^{\ast}\in Q:=\cap_{i=1}^{p}Q_{i}. \label{(CFP problem)}%
\end{equation}
If $Q\ne \emptyset$, the CFP is said to be \emph{consistent}. Consequently, it is required to solve the system of
convex inequalities%
\begin{equation}
\text{ }f_{i}(z)\leq0,\text{ \ \ }i=1,2,\ldots,p. \label{CFP inequalities}%
\end{equation}

The context of convex inequalities gives rise to the realm of subdifferential calculus, in which the definitions of {subgradients} and {subdifferentials} play a fundamental role.
Given the convex function $f_{i}:\mathbb R^{n}\rightarrow \mathbb R$, a vector $t\in\mathbb R^n$ is called a \emph{subgradient} of $f_i$ at point $z_0$ if
\begin{equation}\label{subgrad}
    f_i(z)-f_i(z_0)\ge\langle t,z-z_0\rangle
\end{equation}
The subdifferential of $f_i$ at $z_0$, denoted by $\partial f_i(z_0)$, is the non-empty compact convex set
\begin{equation}\label{subdif}
    \partial f_i(z_0) :=\left\{t: f_i(z)-f_i(z_0)\ge\langle t,z-z_0\rangle,\,\forall z\right\}
\end{equation}
When $f_i$ is differntiable at $z_0$, the subgradient becomes a gradient, $t=\nabla f_i(z_0)$.

\subsection{Subgradient Projections\label{sect:steering}}

Subgradient projections have been incorporated in iterative algorithms for the
solution of CFPs. They can be roughly categorized into two main categories, \emph{sequential} and \emph{simultaneous}. The {cyclic subgradient projections} (CSP) method for the
CFP, which is a sequential subgradient projections algorithm, was developed by Censor and Lent \cite{lc82} and is summarized in Algorithm \ref{algo:CSP}.

\setcounter{theorem}{0}
\begin{algorithm}
\caption{The method of cyclic subgradient projections (CSP)}

\textbf{Initialization}: $z^{0}\in \mathbb R^{n}$ is arbitrary.

\textbf{Iterative step}: Given $z^{k},$ calculate the next iterate $z^{k+1}$
by
\begin{equation}
z^{k+1}=\left\{
\begin{array}
[c]{ll}%
z^{k}-\alpha^{k}\frac{\displaystyle f_{i(k)}(z^{k})}{\displaystyle \parallel
\;t^{k}\parallel^{2}_2}t^{k}, & \mathrm{if}\;\;f_{i(k)}(z^{k})>0,\\
z^{k}, & \mathrm{if}\;\;f_{i(k)}(z^{k})\leq0,
\end{array}
\right.  \label{eqn:CSP}%
\end{equation}
where $t^{k}\in\partial f_{i(k)}(z^{k})$ is a subgradient of $f_{i(k)}$ at the
point $z^{k}$, and the relaxation parameters $\{\alpha^{k}\}_{k=0}^{\infty}$
are confined to an interval $\epsilon_{1}\leq\alpha^{k}\leq2-\epsilon_{2}$,
for all $k\geq0$, with some, arbitrarily small, $\epsilon_{1},\epsilon_{2}>0.$

\textbf{Constraint-Index Control}: 
The sequence $\{i(k)\}_{k=0}^{\infty}$ is cyclic, that is, $i(k)=k(\mod p)+1$ for all $k\geq 0$.
\label{algo:CSP}
\end{algorithm}


A convergence result for the CSP method in the consistent case was provided in \cite[Chapter 5]{CZ97}, where it was shown that
if the functions $f_i(z)$ are continuous and convex on $\mathbb R^n$ $\forall i$; $Q:=\cap_{i=1}^{p}Q_{i}\ne\emptyset$; and
the subgradient is uniformly bounded, then any sequence $\{z^k\}$ produced by
Algorithm \ref{algo:CSP} converges to a solution of the CFP, i.~e., $z^k\rightarrow z^*$ as $ k\rightarrow\infty$.
The convergence proof in \cite[Chapter 5]{CZ97} is based on the concept of {Fej\'{e}r monotonicity}: A sequence $\{z^k\}_{k=0}^{\infty}$ is \emph{Fej\'{e}r monotone} with respect to some fixed set $Q\subseteq \mathbb R^n$, if $\forall z\in Q$,
\begin{equation}\label{fejer}
    \|z^{k+1}-z\|_2\le   \|z^{k}-z\|_2,\,\forall k\ge 0
\end{equation}

Sequential projection
methods for solving CFPs usually have simultaneous counterparts. The
simultaneous subgradient projections (SSP) method \cite{moledo,dossantos} is a simultaneous variant of the CSP, and is given in Algorithm \ref{alg:ssp}.

\begin{algorithm}
\caption{The method of simultaneous subgradient projections (SSP)}

\textbf{Initialization}: $z^{0}\in \mathbb R^{n}$ is arbitrary.

\textbf{Iterative step}: \begin{enumerate}[i.] \item Given $z^{k},$ calculate, for all $i\in
I=\{1,2,\ldots,p\},$ intermediate iterates $\zeta^{k+1,i}$ by
\begin{equation}
\zeta^{k+1,i}=\left\{
\begin{array}
[c]{ll}%
z^{k}-\alpha_{k}\frac{\displaystyle f_{i}(x^{k})}{\displaystyle\parallel
\;t^{k}\parallel^{2}}t^{k}, & \mathrm{if}\;\;f_{i}(z^{k})>0,\\
z^{k}, & \mathrm{if}\;\;f_{i}(z^{k})\leq0,
\end{array}
\right.
\end{equation}
where $t^{k}\in\partial f_{i}(z^{k})$ is a subgradient of $f_{i}$ at the point
$z^{k}$, and the relaxation parameters $\{\alpha_{k}\}_{k=0}^{\infty}$ are
confined to an interval $\epsilon_{1}\leq\alpha_{k}\leq2-\epsilon_{2}$, for
all $k\geq0$, with some, arbitrarily small, $\epsilon_{1},\epsilon_{2}>0.$

\item Calculate the next iterate $z^{k+1}$ by%
\begin{equation}
z^{k+1}=\sum_{i=1}^{p}w_{i}\zeta^{k+1,i}%
\end{equation}
where $w_{i}$ are fixed, user-chosen, positive weights with $\sum_{i=1}%
^{p}w_{i}=1.$
\end{enumerate}
\label{alg:ssp}
\end{algorithm}

The convergence analysis for this algorithm is available only for
consistent ($Q\neq\emptyset$) CFPs, see \cite{dossantos, moledo}.

\subsection{CSP and SSP for Compressed Sensing}

Based on Eqs. \eqref{eq:c1} and \eqref{eq:c2}, we may formulate the following CS problem: Given a real-valued $m\times n$ matrix $H$, a measurement vector $y \in \mathbb R^m$ and a sparse (or compressible) vector $x\in\mathbb R^n$,
\begin{equation}\label{new1}
    \text{find }  x^\ast\in\mathbb R^n \;\text{ s.~t. }\; y=Hx^\ast,\,\|x^\ast\|_1\le\epsilon
\end{equation}

Eq.~\eqref{new1} can be translated into the language of CFP. To do this, we define $m$ closed convex sets $Q_{1},Q_{2},\cdots,Q_{m}\subseteq \mathbb R^{n}$ expressed as%
\begin{equation}
Q_{i}=\left\{  x\in \mathbb R^{n}\mid y_{i}=\langle h_i,\,x\rangle\right\}  , \label{new2}%
\end{equation}
where $h_i$ indicates the $i$-th row of the matrix $H$. In this case, at a given iteration $k$, $t^k=\nabla\langle h_i,\,x\rangle=h_i$. An additional closed convex set $Q_{m+1}\subseteq \mathbb R^{n}$ is defined as
\begin{equation}\label{new3}
    Q_{m+1} = \left\{ x \in\mathbb R^n \mid \|x\|_1-\epsilon \le 0\right\}
\end{equation}
For the latter convex set, we note that a subgradient at the origin can be chosen so that
\begin{equation}\label{ones}
t^k\in\partial \|x\|_1\mid_{x=0}=1_{1\times n}
\end{equation}
where $1_{1\times n}$ is an $n$-dimensional vector of unit entries. It can be easily verified that the subgradient chosen in Eq.~\eqref{ones} satisfies the
subgradient definition of Eq.~\eqref{subgrad}. This subgradient may be generalized $\forall x\in\mathbb R^n$ by adopting the convention
\begin{equation}\label{sign}
    \sign(x_i)=\left\{
                        \begin{array}{c}
                          1,\quad x_i\ge 0 \\
                          -1,\quad x_i< 0 \\
                        \end{array}
     \right.
\end{equation}
and writing, at some iteration $k$,
\begin{equation}\label{sign1}
    t^k=\sign(x^k),\quad\|t^k\|_2^2=n,\,\forall k.
\end{equation}

The next stage is to implement the CSP algorithm in order to find a feasible solution $x^{\ast}\in Q:=\cap_{i=1}^{m+1}Q_{i}$ for the CFP formulation of the CS problem \eqref{new1}. Following the recipe of Algorithm \ref{algo:CSP}, using the relationships \eqref{ones}, \eqref{sign} and \eqref{sign1}, such an algorithm is formulated as follows:

\begin{algorithm}
\caption{Cyclic subgradient projections for compressed sensing (CSP-CS)}

\textbf{Initialization}: $\hat x^{0}\in \mathbb R^{n}$ is arbitrary.

\textbf{Constraint-Index Control}: Set $i(k)=k(\mod p)+1$ for all $k\geq 0$, where $p=m+1$.

\textbf{Iterative step}: For $1\le i \le m$, calculate $\hat x^{k+1}$
by
\begin{equation}
\hat x^{k+1}=\hat x^{k}-\alpha^{k}\frac{\displaystyle \langle h_{i},\,\hat x^k\rangle-y_i }{\displaystyle \parallel
\;h_i\parallel^{2}_2}h_{i}
\label{new4}%
\end{equation}
with the relaxation parameters $\{\alpha^{k}\}_{k=0}^{\infty}$ confined to the open interval $(0,\,2)$.

For $i=p=m+1$, calculate $\hat x^{k+1}$
by
\begin{equation}
\hat x^{k+1}= \hat x^{k}-\lambda^{k}\left[\frac{\displaystyle \|\hat x^k\|_1-\epsilon }{\displaystyle n }\sign(\hat x^k)\right]
\mathcal{H}[\|\hat x^k\|_1-\epsilon]
\label{new5}%
\end{equation}
where $\mathcal{H}[x]$ denotes the standard step function (i.e., $\mathcal{H}[x]=1$ for $x > 0$ and $\mathcal{H}[x]\le 0$ otherwise),
and $\{\lambda^{k}\}_{k=0}^{\infty}$ are confined to the open interval $(0,\,2)$.
\label{alg:cspcs}
\end{algorithm}
%
%
%

An application of Algorithm \ref{alg:ssp} for the CS problem \eqref{new1}, termed SSP-CS, can be done in a similar manner to Algorithm \ref{alg:cspcs}, and is omitted here for the sake of conciseness.

Note that the inactive step of Algorithms \ref{algo:CSP} and \ref{alg:ssp}, i.~e., when the current iterate remains unchanged, is unnecessary if the constraints are linear equations (cf.~\cite[Chapter 6]{CZ97}). This fact helps increasing the computational efficiency.  In the same context, it is worth noting that without using the 1-norm constraint implemented in Eq.~\eqref{new5}, the recovery of $x$ from the linear set of equations
$y=Hx$ is possible, under certain regularity conditions, using the well-known algorithm of Kaczmarz\cite[Chapter 6]{CZ97}. In this case, the obtained solution
$x^\ast$ would minimize $\|x\|_2$. However, adding the 1-norm constraint is key to a successful recovery when $x$ is sparse (or compressible).

The convergence of Algorithms \ref{alg:cspcs} and its SSP variant to some feasible solution $x^\ast$ is guaranteed, provided that the
CFP is consistent (i.~e., the solution set is non-empty). However, for real-world, large-scale problems, one cannot determine \emph{a-priori} whether a given value of $\epsilon$ renders the CFP consistent; this is essentially a matter of tuning. The behavior of the subgradient projections algorithms for the inconsistent case was investigated in Ref.~\cite{butnariu}, where it was shown that simultaneous subgradient projections methods usually yield better convergence in the inconsistent case. However,  simultaneous projection methods tend to converge slower than sequential methods; it is thus a good idea to combine these two methods in order to fight instabilities and obtain fast convergence. This issue is discussed in the next section.

\section{Practical Implementation}

In this section we discuss several issues concerning the practical implementation of the CSP-CS and SSP-CS algorithms.

\subsection{Stopping Criteria}

The CSP-CS and SSP-CS routines are terminated when either some predetermined maximal number of iterations is exceeded, or
when there is no significant difference between two consecutive estimates, viz.
$\parallel \hat x^{k+1} - \hat x^k \parallel_2 \le \gamma$ for some small $\gamma > 0$.

\subsection{Refinements}
\label{sec:cspvar}

As mentioned in the previous section, in some cases the set of constraints may not be consistent due to an improper setting of $\epsilon$, which in turn implies that
the convergence of the CSP method is not guaranteed. Typically, in such scenarios the estimated parameters persistently
fluctuate around some nominal value. The reconstructed signal itself can be highly
accurate; however, its entries do not attain a fixed value. This problem can be alleviated by incorporating an
additional refinement stage. In this work, we propose two such refinement approaches, each of which relies on a different technique for
producing refined estimates.

\subsubsection{Gauss-CSP}

After a given number of CSP-CS iterations, the support of the unknown signal can be readily approximated based
on the magnitudes of the estimated elements. As soon as the support is given, the signal can be accurately reconstructed
following a simple least squares (LS) procedure. This approach is inspired by the strategy adopted in \cite{dantzig:06}
for correcting the bias inherent to the Dantzig selector (the corrected scheme was referred to in \cite{dantzig:06} as the Gauss-Dantizg selector). This technique, applied to the CSP-CS algorithm, will be referred to as the \emph{Gauss-CSP}. This variant of CSP-CS is summerized in
Algorithm~\ref{alg:gauss-csp}.
\begin{algorithm}
\caption{Gauss-CSP}

\textbf{CSP-CS Stage}: Perform several CSP-CS iterations based on Algorithm \ref{alg:cspcs} until a stopping criteria is reached.

\textbf{LS Stage}: Set $N < m$ as the maximal support size of the obtained estimate $\hat x^k$. Define a set of indices $G:=\{j_i\}^N_{i=1}$
of the most significant entries in $\hat x^k$ in terms of magnitude. Solve
\begin{equation}
\beta = \left(\hat H^T \hat H\right)^{-1} \hat H^T y
\end{equation}
where $\hat H \in \mathbb{R}^{m \times N}$ is composed of the columns of $H$ corresponding to the indices in $G$.
Set $\hat x^k_l = 0$, $\forall l \notin G$ and $\hat x^k_{j_i} = \beta_i$, $i=1,\ldots,N$.
\label{alg:gauss-csp}
\end{algorithm}

A slightly different implementation of the Gauss-CSP, referred to as \emph{alternating Guass-CSP}, can dramatically improve the convergence properties of the CSP method.
This consists of incorporating the LS stage directly into the CSP algorithm in an alternating manner. This variant of the Gauss-CSP
typically converges much faster than any other alternative. Following this approach, the LS routine is executed
whenever the cyclic index $i$ in Algorithm~\ref{alg:cspcs} reaches an arbitrary predetermined value.

\subsubsection{CSP-SSP}

Normally, running a few SSP iterations subsequent to the termination of the CSP routine improves the accuracy of recovery. This stems from the improved behavior of simultaneous subgradient projections method in the inconsistent case, discussed in the previous section.

\subsection{Block Processing}

In some programming environments, it might be more computationally efficient to process a group of $l$ observations
$y_i = \langle h_i ,x\rangle$, $i=j,\ldots, j+l$ at each iteration rather than a single one.
Our own implementation of the CSP-CS exploits this simple idea for alleviating the workload on the \texttt{MATLAB}\textsuperscript{\textregistered} interpreter, which may become prohibitively slow when loops are involved. This approach does not require any considerable modification of the
original algorithm. In practice, this is accomplished by running the cyclic index from 1 to $p/l+1$ while
repeating the CSP update stage \eqref{new4} $l$ times at each iteration (one update for each observation in the block).


\section{Illustrative Examples}

In this section, the new convex feasibility programming algorithms are assessed based on an extensive comparison with some of the commonly-used
methods for CS. The algorithms considered herein consist of the homotopy method LARS~\cite{lars:04},
the two greedy algorithms, OMP~\cite{omp} and BP~\cite{bp:98}, the recently-introduced gradient projection (GP)-based method of \cite{gp}, and
the Bayesian CS (BCS) of \cite{bcs} (the links for the \texttt{MATLAB}\textsuperscript{\textregistered} implementations of the various methods used
here are provided in the Appendix).
In order to highlight the weaknesses and virtues of the various methods, we examine
here both synthetic and realistic scenarios, associated with the two signal types frequently encountered in CS applications: Sparse and compressible (nearly sparse in the sense of \eqref{eq:sd}).
An indicator for the difficulty of recovery is assigned for each
problem based on its unique settings (dimension and sparseness degree). This measure, termed here the recovery index, is derived
from the relation $m \ge c (s \log n)$~\cite{cs:06} (for some $c > 0$) as
\begin{equation}
\label{eq:ri}
(s/m) \log n
\end{equation}
This index essentially refers to the probability of recovery assuming the sensing matrix obeys the RIP up to a certain
sparseness degree. As it was already pointed out in \cite{cs:06}, as this measure increases an exact recovery becomes less probable.

Throughout this section, the signal reconstruction error is computed as
\begin{equation}
e := \sqrt{\frac{\parallel x - \hat x \parallel_2^2}{d(x)}}
\end{equation}
where the normalizing term $d(x)$ is set according to the signal type,
\begin{equation}
d(x):= \left \{ \begin{array}{ll}
\sum^n_{i=1} \min(x_i^2, \sigma^2), & \text{for sparse $x$}\\
\parallel x \parallel_2^2, & \text{for compressible $x$}
\end{array} \right.
\end{equation}
with $\sigma$ being the standard deviation of the observation noise. The above formulation consolidates both the \textit{ideal}
and the \textit{normalized} recovery measures that are used for assessing the reconstruction accuracy in \cite{dantzig:06} and
\cite{bcs}, respectively. The mean reconstruction error is computed in a similar fashion over $N$ runs via $\sqrt{N^{-1} \sum_i e^2(i)}$, where $e(i)$ denotes the error in the $i$th run.

\subsection{Least Squares Augmentation}

The approach underlying the Gauss-Dantzig~\cite{dantzig:06} and the Gauss-CSP in Algorithm~\ref{alg:gauss-csp},
which essentially consists of incorporating an additional LS stage for refining the obtained estimates, can be
easily applied for any other CS method. This poses a question as to what should be a fair comparison
when considering the Gauss-CSP variant. Intuitively, one could think that the relative improvement of the LS stage
as described in Algorithm~\ref{alg:gauss-csp} would be identical irrespectively of the CS method, and therefore such variants
should not be compared at all. This statement, however, turns out to be incorrect, as shown by our experiments.
In order to maintain our observations as equitable as possible in these circumstances, our
comparison involves, in addition, an LS-augmented variant of each of the aforementioned CS techniques.

\subsection{Synthetic Example}

In the first scenario, the various methods are applied for the recovery of a sparse signal from noisy observations.
The sensing matrix $H \in \mathbb{R}^{m \times n}$ used here consists of normalized random entries sampled from a zero-mean
Gaussian distribution with a standard deviation of $1/\sqrt{m}$. This type of random construction has been shown to obey the RIP up to a reasonable
sparseness level (see \cite{candes:06}). We examine the recovery performance of the CS algorithms in various settings consisting of
different problem dimensions, ranging from 512x1024 (small) through 1024x2048 and 2048x4096 (medium) to 3072x6144 (large), and
different sparseness levels. The original signal $x$ is composed of only few nonzero random elements, which are
uniformly sampled over $[-1,1]$, and of which the indices are randomly picked between 1 and $n$.
In all runs, the measurement noise standard deviation is set as $0.01$. The various algorithms' tuning parameters
are taken as those that seemed to minimize the recovery error based on tuning runs. The CSP-CS relaxation parameters are set as
\begin{equation}
\alpha^k = 1.8, \;\;\; \epsilon=10^{-4}, \;\;\; \frac{\lambda^k}{n} = \left \{
\begin{array}{ll}
70^{-2}, & \text{If } k \le 2000\\
100^{-2}(1+k/10^4)^{-1}, & \text{Otherwise}
\end{array}
\right.
\end{equation}
which seemed to be the best values with respect to accuracy and convergence time.
The CSP routine is terminated when either the number of iterations exceeds 5000 or when the normed difference
between two consecutive estimates, $\parallel \hat x_{k+1} - \hat x_k \parallel_2$, drops below some $\gamma$
where $\gamma=0.01$ for the low dimensional problem and $\gamma=0.5$ otherwise.

Both the OMP and the LS refinement stage assume a maximal support size of the reconstructed signal. In all our experiments
we have chosen this parameter as $1.5s$ where $s$ is the actual support size (sparseness level) of $x$.

\begin{figure}[htb]
\centering
\subfloat[Convergence Time]{
\includegraphics[width=0.50\textwidth]{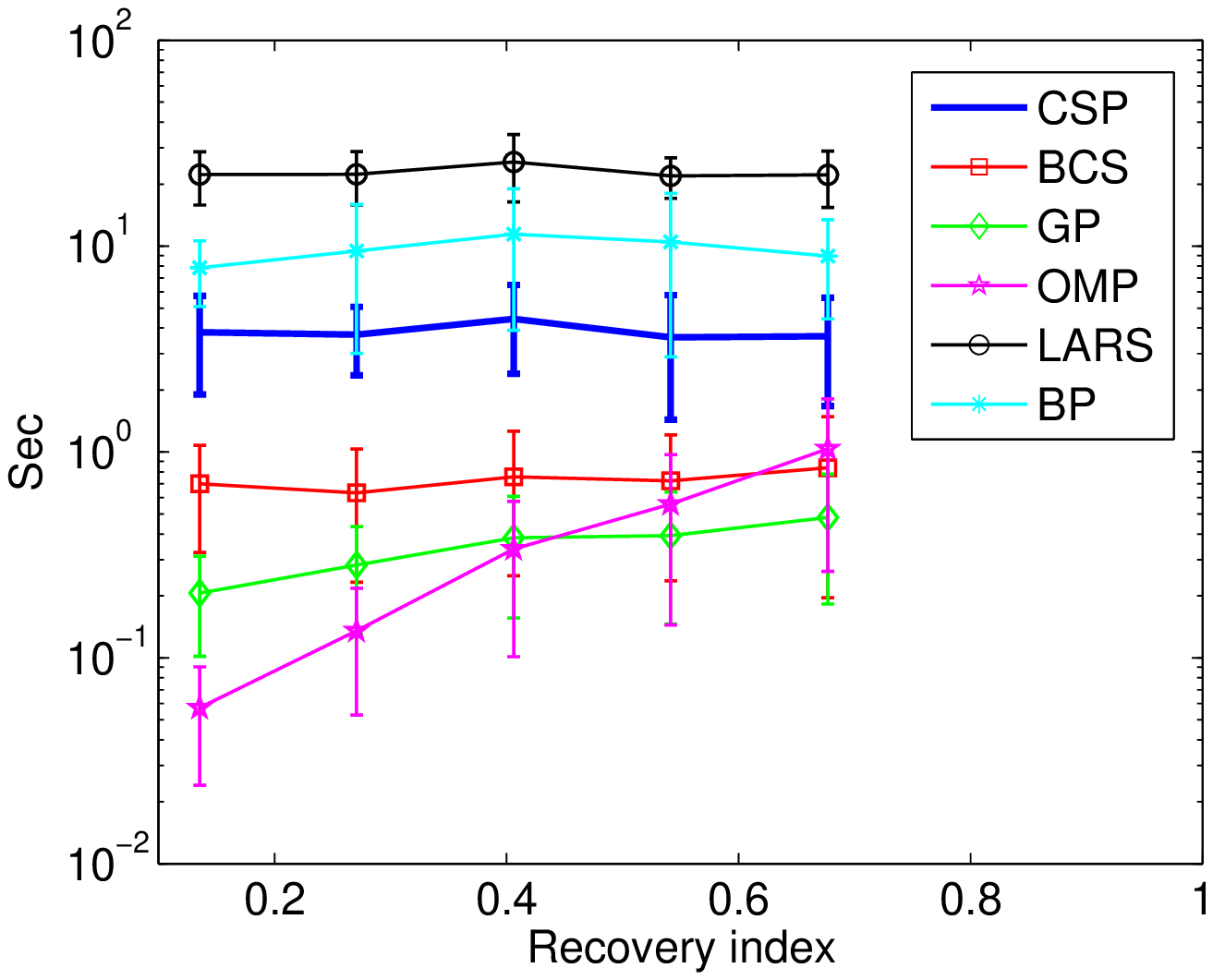}}
\subfloat[Ideal normalized error]{
\includegraphics[width=0.50\textwidth]{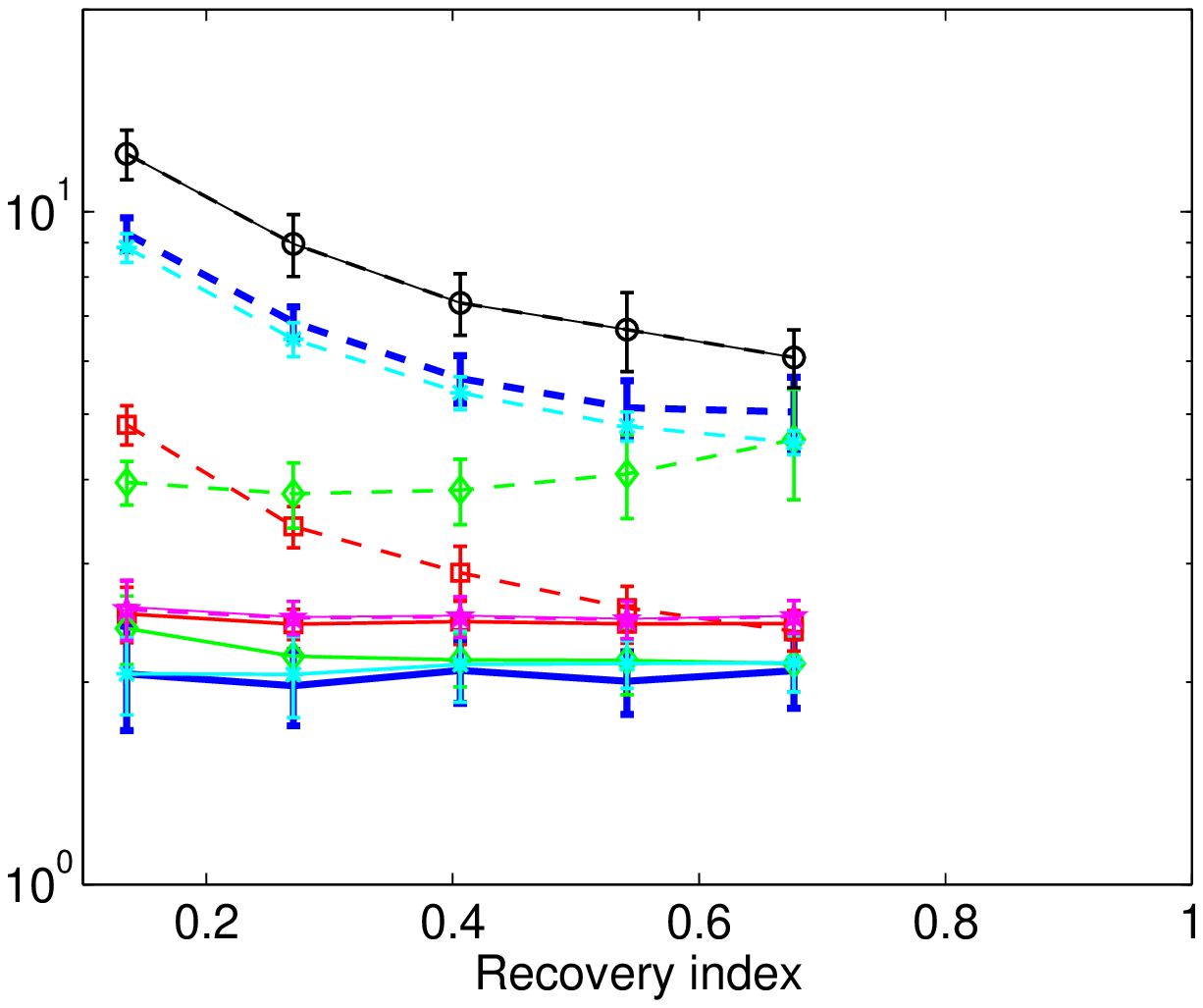}}

\caption{The average performance (over 50 Monte Carlo runs) of the CS methods (including the computationally excessive ones, the LARS and the BP)
for problem dimension 512x1024. The performance of the LS-augmented variants is depicted via solid lines.}
\label{fig:small_sparse}
\end{figure}

The averaged performance over 50 Monte Carlo runs of the various methods in the small-scale scenario is depicted in
Fig.~\ref{fig:small_sparse} for different recovery indices ranging from $0.1$ to $0.7$. This figure shows both the
mean ideal recovery error and the corresponding mean convergence time along with their standard deviations, which are illustrated using error bars. Observing the right panel in this figure reveals a performance hierarchy
that places the CSP as the 2nd worst right after the LARS and just a bit before the BP.
The remaining methods attain higher accuracy in this case.

Nevertheless, the story completely changes
when examining the LS-augmented methods. Here the Gauss-CSP attains the \emph{best accuracy} over the entire range
of sparseness degrees. The advantage of using the LS stage is likewise prominent in all other methods, as it significantly reduces the recovery error.
By observing the left panel of Fig.~\ref{fig:small_sparse}, it can be
easily recognized that the CSP is the 3rd slowest method in this case. The other computationally excessive methods
here are the LARS and the BP. In this small-scale example, the convergence time of all methods excluding those of the greedy OMP
and the GP roughly stays unchanged over the entire range of sparseness levels. The extreme slant of the OMP line
is due to the nature of this algorithm, which tends to become computationally intensive as the assumed maximal support size
increases.

Note that the convergence times of the LS-augmented
methods are omitted here and in the sequel, as they are nearly identical to those of the original unaugmented methods
(the LS stage is implemented using the \texttt{MATLAB}\textsuperscript{\textregistered} pseudo-inverse command \texttt{pinv}, which is extremely fast even in high dimensions).

The above insights are further emphasized in Table~\ref{tab:small_sparse}, which repeats
the values from Fig.~\ref{fig:small_sparse} for two nominal recovery indices.
The bold values in this table correspond to the averaged recovery errors of the LS-augmented methods. Thus,
it can be clearly seen that the Gauss-CSP attains the best accuracy yielding
a mean recovery error of around $2.05$.

\begin{table}[tbh]
\centering
\begin{tabular}{|l|c|c|c|c|c|c|}
\hline\hline
Method & \multicolumn{2}{c|}{$(s/m) \log n=0.1$} & \multicolumn{2}{c|}{$(s/m) \log n=0.6$} & $(s/m) \log n=0.1$ & $(s/m) \log n=0.6$ \\
\hline
LARS & 12.19 & \textbf{12.19} & 6.07  & \textbf{6.07} & 22.28 (sec)& 22.18 (sec)\\
BP   & 8.84  & \textbf{2.05} & 4.54  & \textbf{2.13}  & 7.84 (sec) & 8.93 (sec)\\
OMP  & 2.56  & \textbf{2.58} & 2.50  & \textbf{2.50}  & 0.05 (sec) & 1.03 (sec)\\
BCS  & 4.82  & \textbf{2.52} & 2.37  & \textbf{2.44}  & 0.70 (sec) & 0.83 (sec)\\
GP   & 3.95  & \textbf{2.40} & 4.58  & \textbf{2.12}  & 0.20 (sec) & 0.48 (sec)\\
CSP  & 9.28  & \textbf{2.05} & 5.04  & \textbf{2.07}  & 3.81 (sec) & 3.64 (sec)\\
\hline\hline
\end{tabular}

 \caption{Recovery of sparse signals. The ideal recovery error (left columns) and convergence time (right columns)
of the various methods for the problem
dimension 512x1024. The bold values correspond to the accuracy of the two-staged LS-augmented variants. Averaged over 50 Monte Carlo runs.}
\label{tab:small_sparse}
\end{table}

\begin{figure}[htb]
\centering
\subfloat[1024x2048]{
\includegraphics[width=0.33\textwidth]{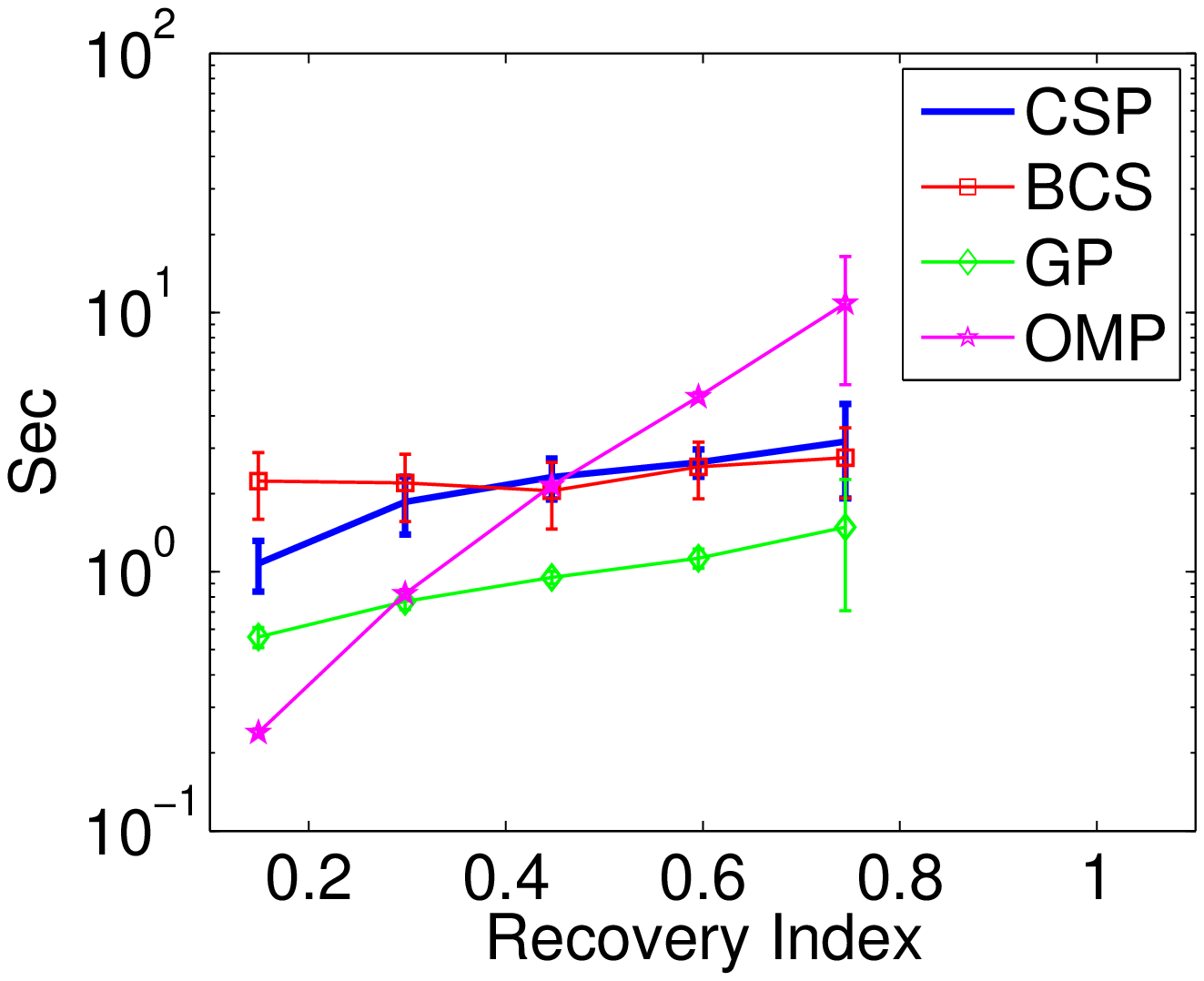}}
\subfloat[2048x4096]{
\includegraphics[width=0.33\textwidth]{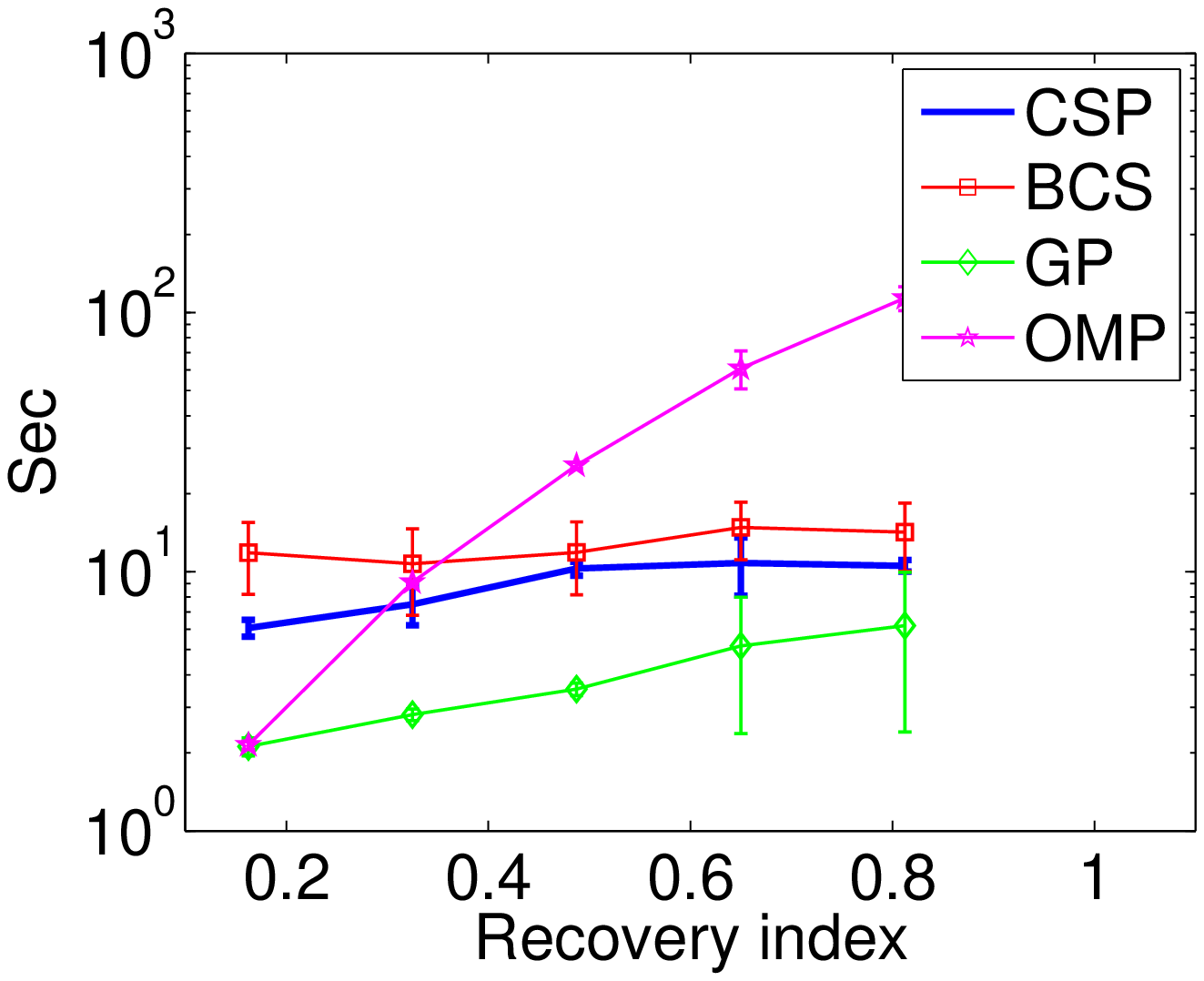}}
\subfloat[3072x6144]{
\includegraphics[width=0.33\textwidth]{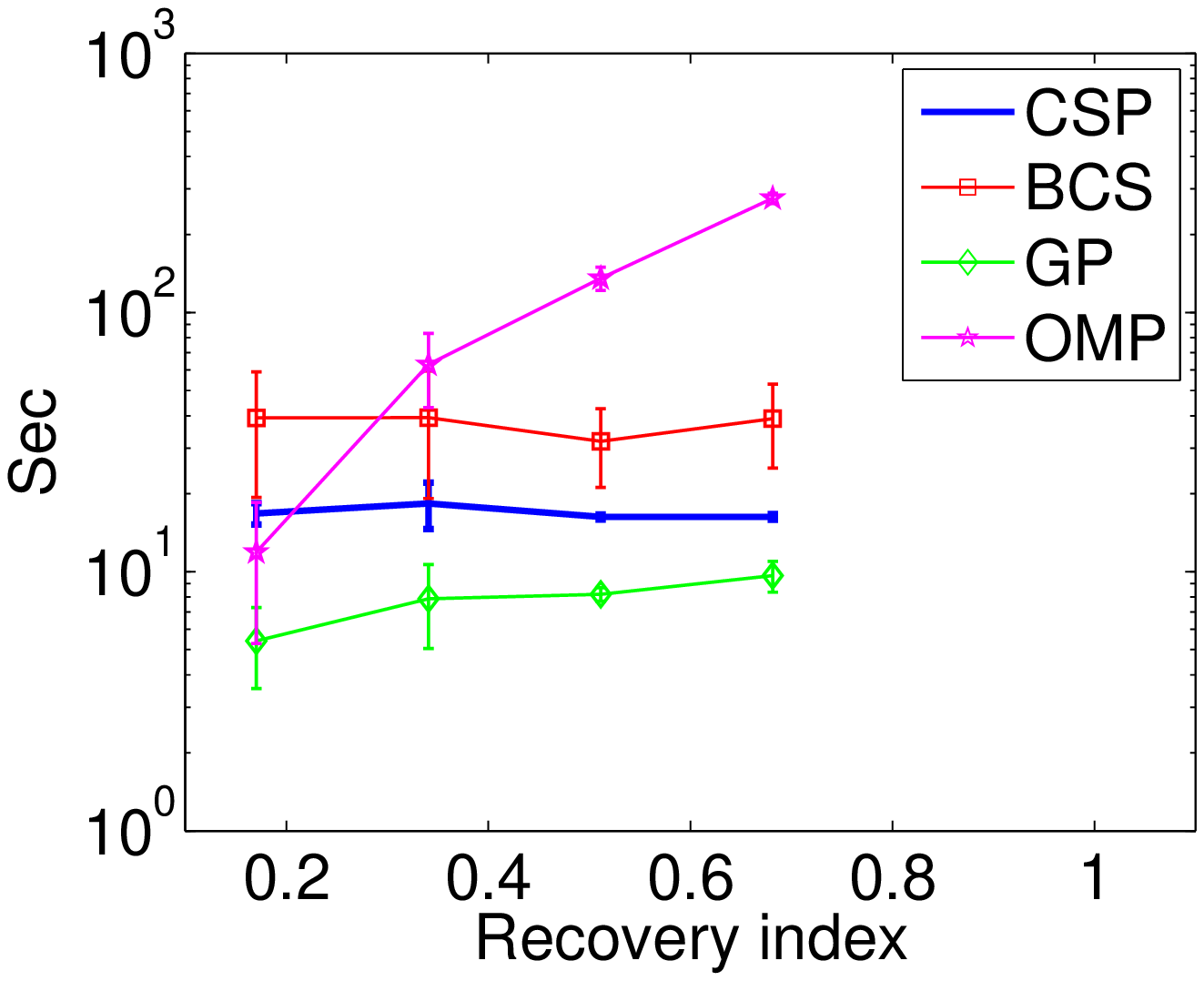}}\\
\subfloat[1024x2048]{
\includegraphics[width=0.33\textwidth]{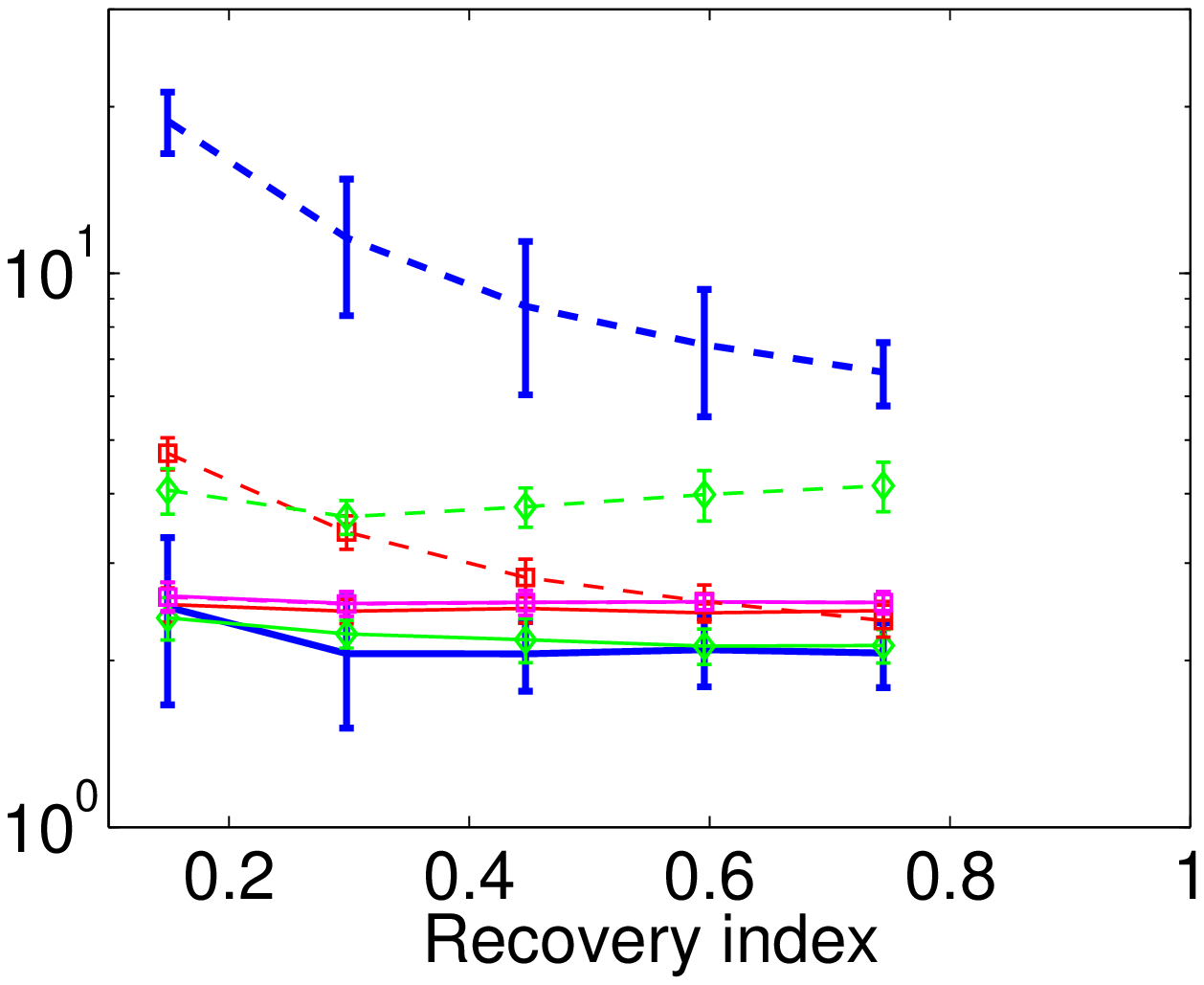}}
\subfloat[2048x4096]{
\includegraphics[width=0.33\textwidth]{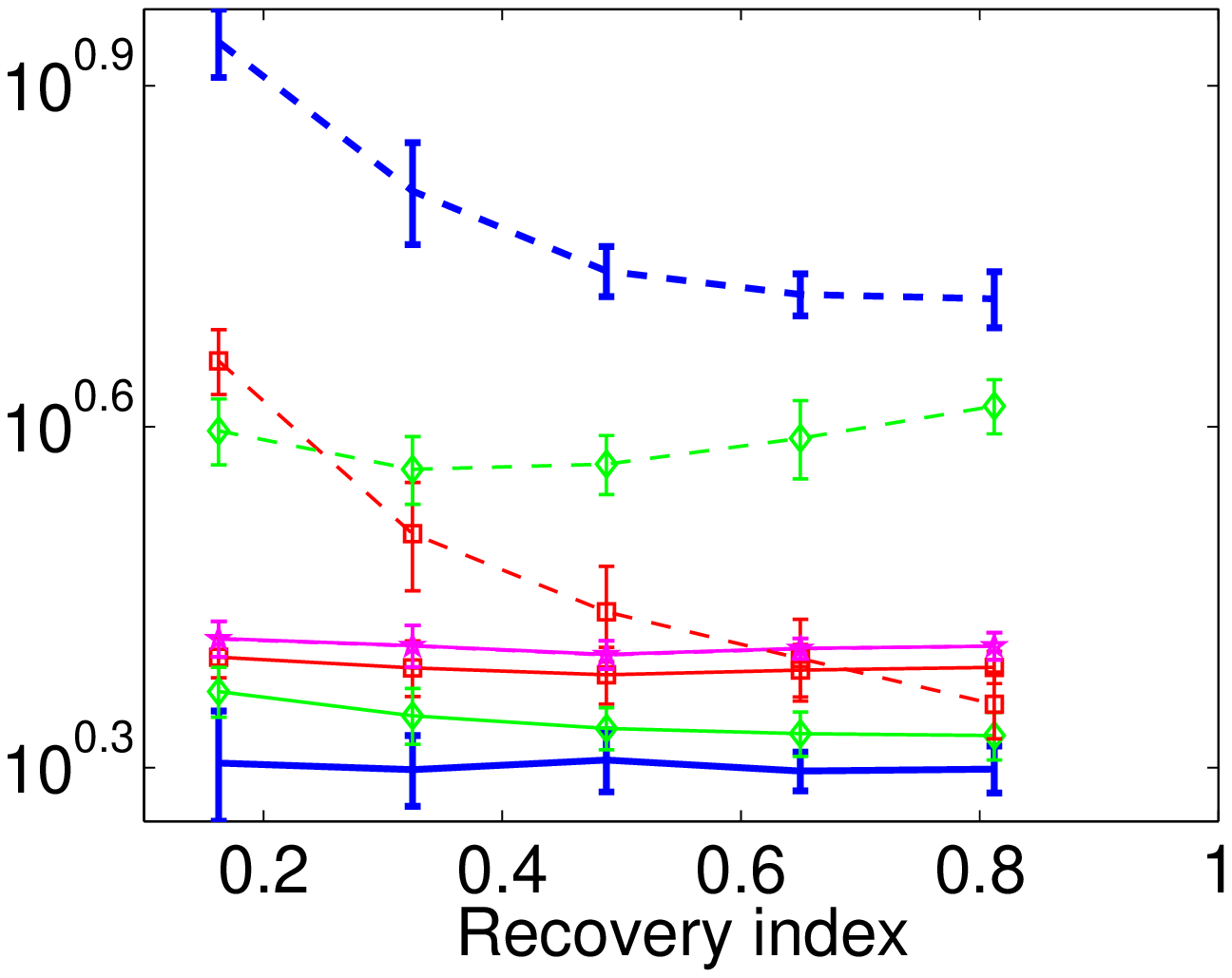}}
\subfloat[3072x6144]{
\includegraphics[width=0.33\textwidth]{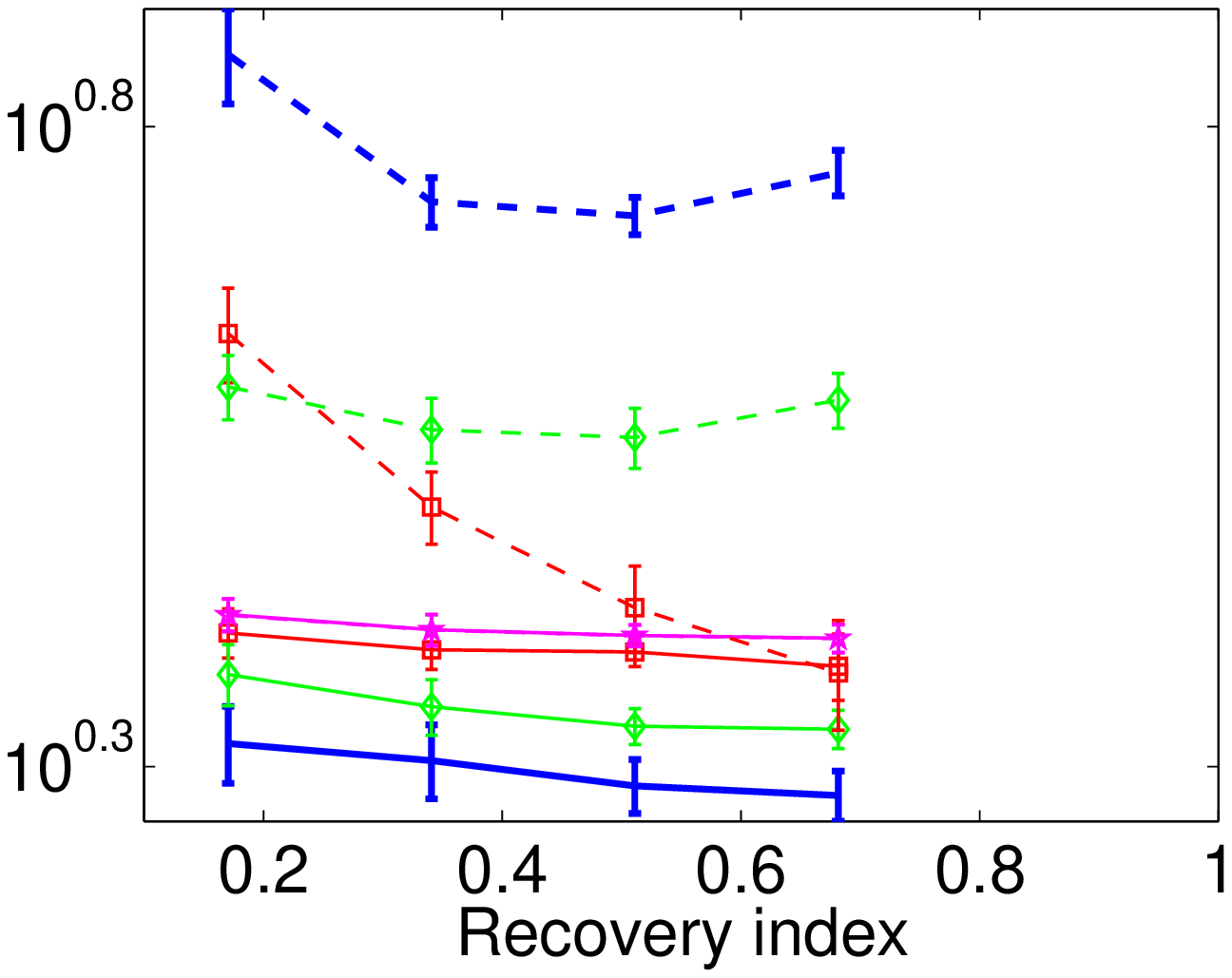}}

\caption{The average performance of the CS algorithms for various problem dimensions and sparseness degrees.
Showing the convergence time (upper panel) and the ideal recovery error (bottom panel).
The computationally excessive methods (LARS and BP) are not shown here.}
\label{fig:ml_sparse}
\end{figure}

The performance of the algorithms in the medium and large scale scenarios is illustrated in Fig.~\ref{fig:ml_sparse}.
Here we have excluded the computationally intensive methods, the LARS and the BP, as their convergence time became
prohibitively long. The performance of the remaining methods for the various problem dimensions is illustrated
via two panels. Thus, the upper panel shows the convergence times for different sparseness levels, whereas the bottom panel shows the corresponding ideal recovery errors averaged over 50 Monte Carlo runs.
As before, the standard deviations from the mean values are depicted using error bars.

By observing the upper panel in this figure, it can be easily recognized that the CSP
method is comparable in speed and even faster than the BCS over the entire range of sparseness levels
in the medium scale scenarios.
A similar conclusion applies when comparing the CSP with the OMP from a certain sparseness degree corresponding to
a recovery index of around $0.4$. The CSP turns out to be \emph{significantly faster} than both the BCS and the OMP
over almost the entire range of sparseness levels in the large scale scenario.

The corresponding recovery errors of the various methods are presented in the bottom panel in this figure.
Thus, it can be recognized that the unaugmented CSP yields a slightly worse accuracy than the other methods (notice the
logarithmic ordinate). The 2nd worse method in terms of accuracy in this case is also the fastest of them all, the GP.
Nevertheless, when looking at the augmented methods it turns out that, as we have already witnessed in the small-scale scenario, the Gauss-CSP is the best method in terms of accuracy. The 2nd best in this case is the LS-augmented GP.

The above insights are supported by Table~\ref{tab:ml_sparse}, which provides the timing and
recovery error values in the large scale scenario for two nominal recovery indices. Thus, it
can be easily seen that the Gauss-CSP outperforms the other methods in terms of accuracy and is also
the 2nd fastest method (after the GP). Another interesting and important detail that stems from both Table~\ref{tab:ml_sparse}
and Fig.~\ref{fig:ml_sparse} is related to the fact that as opposed to the GP, of which the running time is highly sensitive to
the sparseness degree, the CSP computation time remains almost unchanged with respect to this factor.
This observation could, in fact, be expected, as the CSP mechanism does not really distinguish between elements in the support
and those which are not. This renders it highly robust and efficient when applied either to compressible problems or
in such scenarios where the recovery index is relatively large.

\begin{table}[tbh]
\centering
\begin{tabular}{|l|c|c|c|c|c|c|}
\hline\hline
Method & \multicolumn{2}{c|}{$(s/m) \log n=0.1$} & \multicolumn{2}{c|}{$(s/m) \log n=0.6$} & $(s/m) \log n=0.1$ & $(s/m) \log n=0.6$ \\
\hline
OMP  & 2.62  & \textbf{2.62} & 2.52  & \textbf{2.52}  & 11.89 (sec) & 135.52 (sec)\\
BCS  & 4.35  & \textbf{2.53} & 2.65  & \textbf{2.45}  & 39.18 (sec) & 31.84 (sec)\\
GP   & 3.95  & \textbf{2.35} & 3.60  & \textbf{2.14}  & 5.41 (sec) & 8.18 (sec)\\
CSP  & 7.18  & \textbf{2.08} & 5.37  & \textbf{1.92}  & 16.79 (sec) & 16.26 (sec)\\
\hline\hline
\end{tabular}

 \caption{Recovery of sparse signals. The ideal recovery error (left columns) and convergence time (right columns)
of the various methods for the problem dimension 3072x6144. The bold values correspond to the accuracy of the two-staged LS-augmented variants.
Averaged over 50 Monte Carlo runs.}
\label{tab:ml_sparse}
\end{table}

\begin{figure}[htb]
\centering
\includegraphics[width=0.60\textwidth]{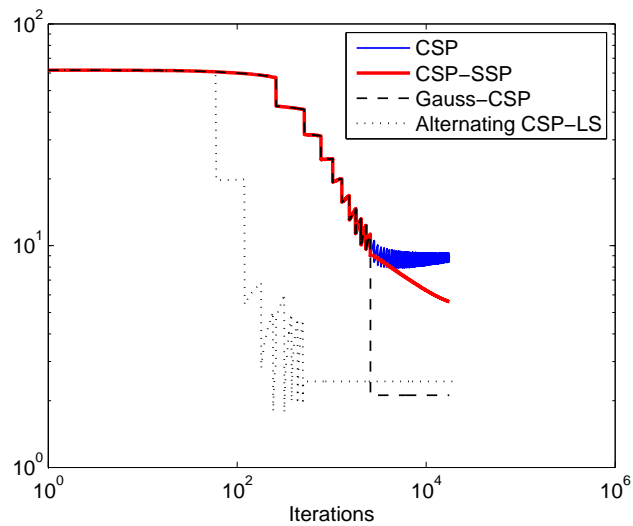}

\caption{The ideal recovery error of the CSP variants. Single run, problem dimension 2048x4096.}
\label{fig:cspvar}
\end{figure}

A comparison of the various CSP implementations that were discussed in Section~\ref{sec:cspvar} is provided in Fig.~\ref{fig:cspvar}.
This figure demonstrates the convergence properties of the methods based on a single run for a problem dimension of 2048x4096.
Thus, it can be easily recognized that the plain CSP attains the worst recovery error, as it begins to fluctuate around some
nominal value. As was pointed out previously, this behavior indicates that the set of constraints in this case is
inconsistent owing to an improper setting of the parameter $\epsilon$. This problem is alleviated in both of the variants,
the CSP-SSP and the Gauss-CSP. Although both these methods outperform the plain CSP, it seems that the best attainable error is achieved
by using the Gauss-CSP, whereas the fastest convergence is obtained by using the alternating CSP-LS scheme.

%
%

\subsection{Realistic (Compressible) Examples}

Compressible signals are of greater practical importance when it comes to real-world applications.
As was previously pointed out, such signals are nearly sparse in the sense that they consist of many
relatively small elements that are not identically zero.
Following this, we consider here two realistic experimental studies involving compressible signals.
The first example, which is conducted in the spirit of the previous synthetic one, consists of constructing the discrete Fourier
transform (DFT) of an undersampled time series. The second experiment, which follows right after, involves the reconstruction of
the famous Shepp-Logan phantom head image that is commonly used for assessing the performance of recovery schemes in tomography.

\subsubsection {Constructing a DFT from Undersampled Data}

In this example we consider a discrete signal in the time domain, which takes the form
\begin{equation}
y_k = \sum^{n_f}_{i=1} \sin(\omega_i t_k)
\end{equation}
where the frequencies $\omega_i$, $i=1,\ldots, n_f$ are uniformly sampled over $[1,10\pi]$.
Let $x \in \mathbb{R}^n$ be the DFT of $y_k$ over the discrete times $k=1,\ldots,n$, that is
\begin{equation}
x_k = 1/\sqrt{n}  \sum^n_{j=1} y_j \exp\left(- 2\pi (j-1)(k-1)i/n\right)
\end{equation}
which can be written compactly as $x = F y$ with $F$ being the unitary DFT matrix. Now, suppose that we wish
to reconstruct $x$ from a randomly sampled set of points $\{y_j\}^m_{j=1}$ for which $m << n$. In other words,
we attempt to compose the DFT $x$ from an undersampled time series. This is a classical CS problem, for which
the sensing matrix $H$ is given as a partial inverse DFT matrix consisting of only $m$ randomly-picked
rows. Thus,
\begin{equation}
y = H x
\end{equation}
where $H$ is composed of $m$ randomly picked rows from $F^{*}$, corresponding to the time points in $y$.
The vector $x$ itself is expected to be compressible, essentially comprised of decaying coefficients in the vicinity
of those which are associated with the underlying frequencies $\omega_i$, $i=1,\ldots, n_f$.

Similarly to the synthetic case, we apply the various CS methods (excluding the computationally excessive ones)
for different problem dimensions and sparseness levels. As distinct from the previous example, here the recovery
index \eqref{eq:ri} is computed based on the \emph{effective} sparseness measure \eqref{eq:sd}
with $\varepsilon=0.05 \max_i |x_i|$. The CS algorithms' tuning parameters are chosen to maximize accuracy based on tuning runs. The CSP relaxation variables and termination conditions remain unchanged.

\begin{figure}[htb]
\centering
\subfloat[512x1024]{
\includegraphics[width=0.33\textwidth]{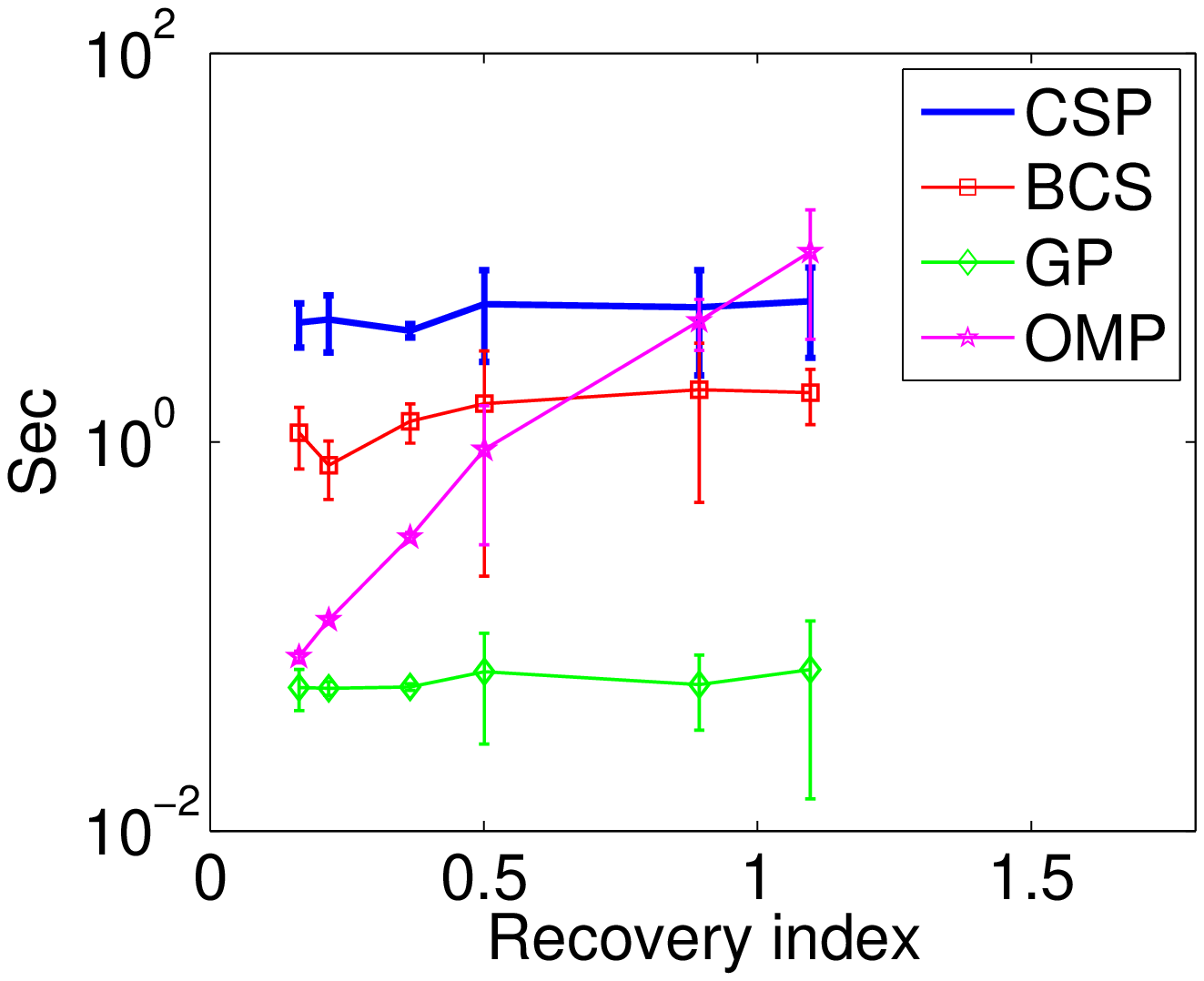}}
\subfloat[1024x2048]{
\includegraphics[width=0.33\textwidth]{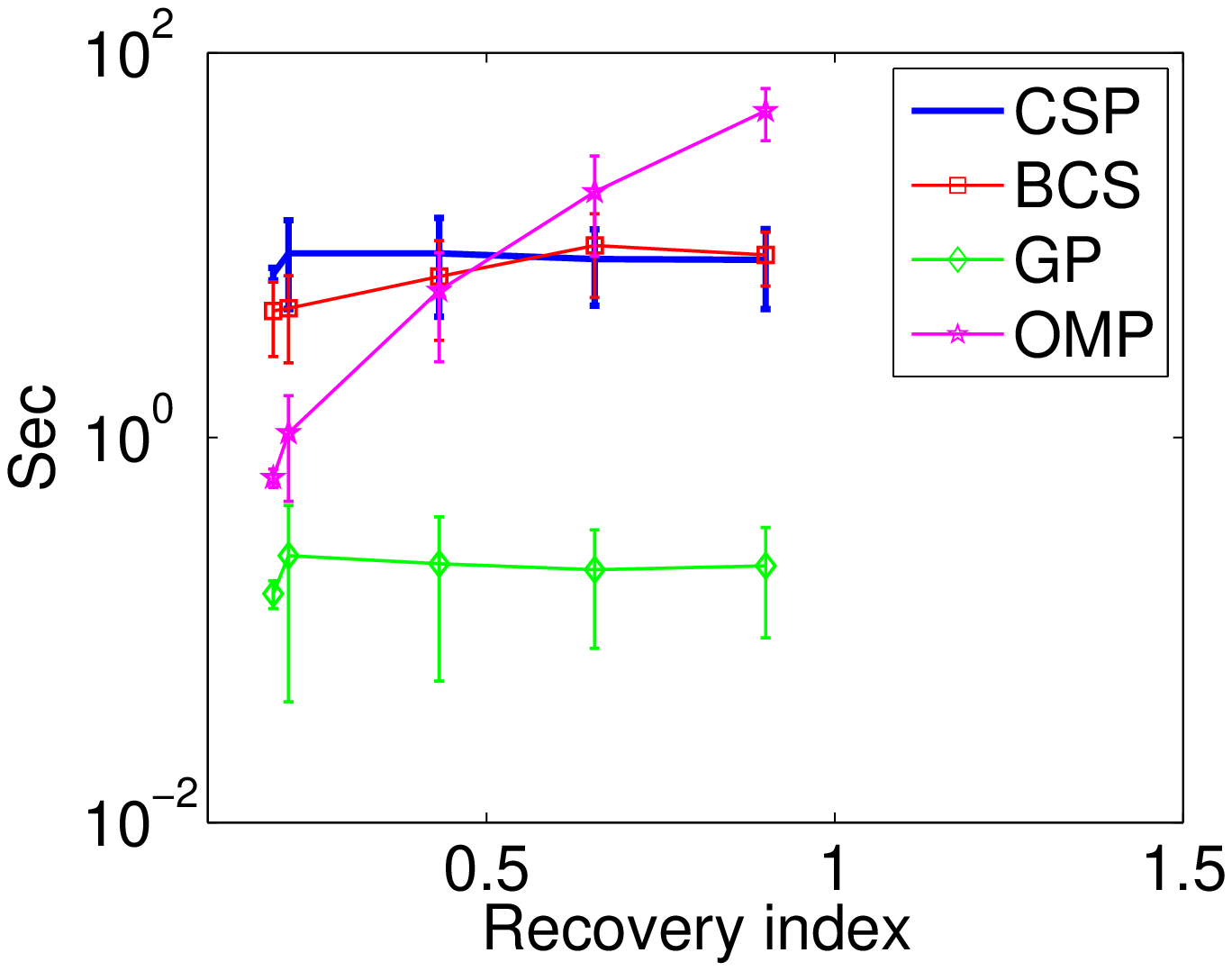}}
\subfloat[2048x4096]{
\includegraphics[width=0.33\textwidth]{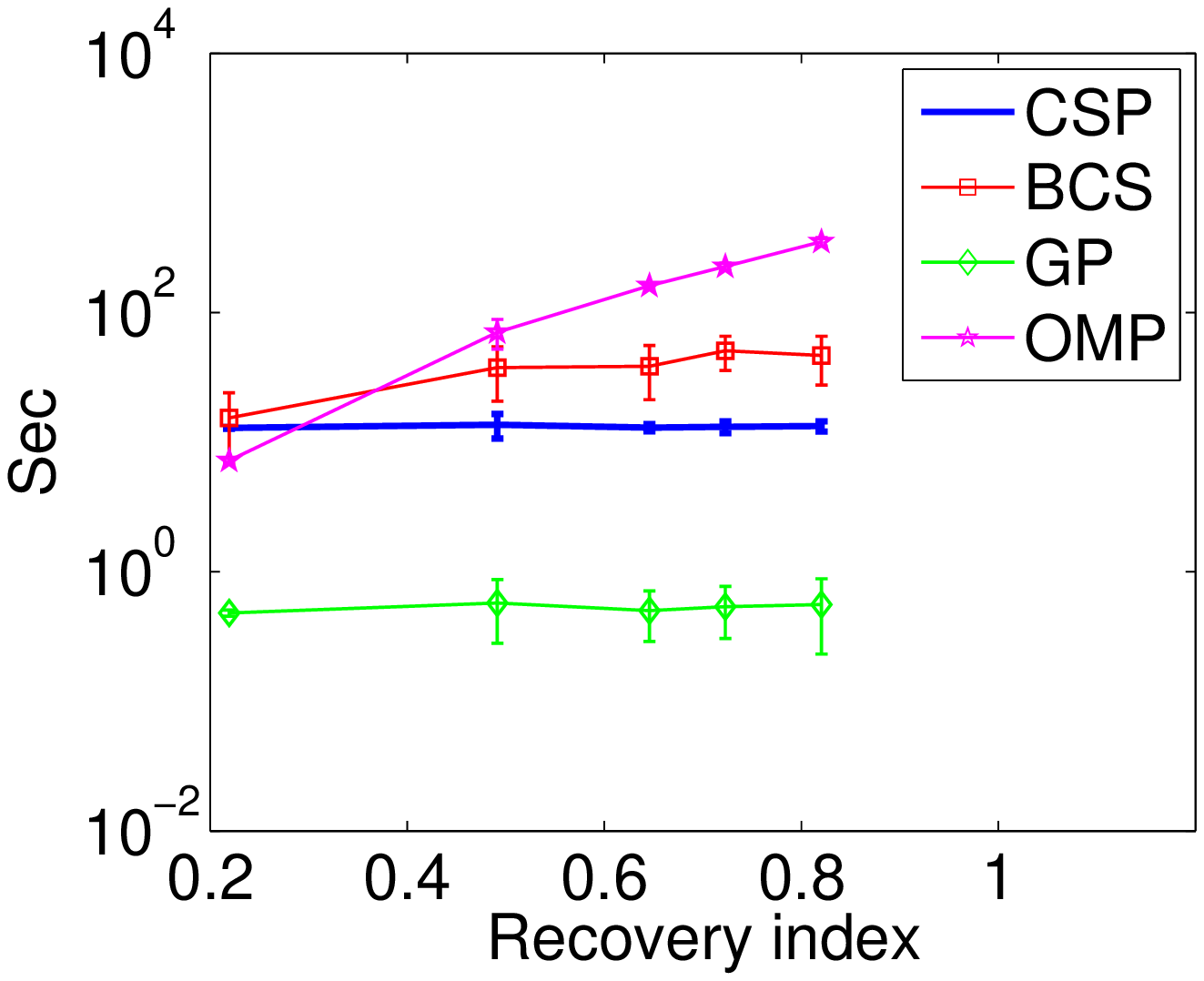}}\\
\subfloat[512x1024]{
\includegraphics[width=0.33\textwidth]{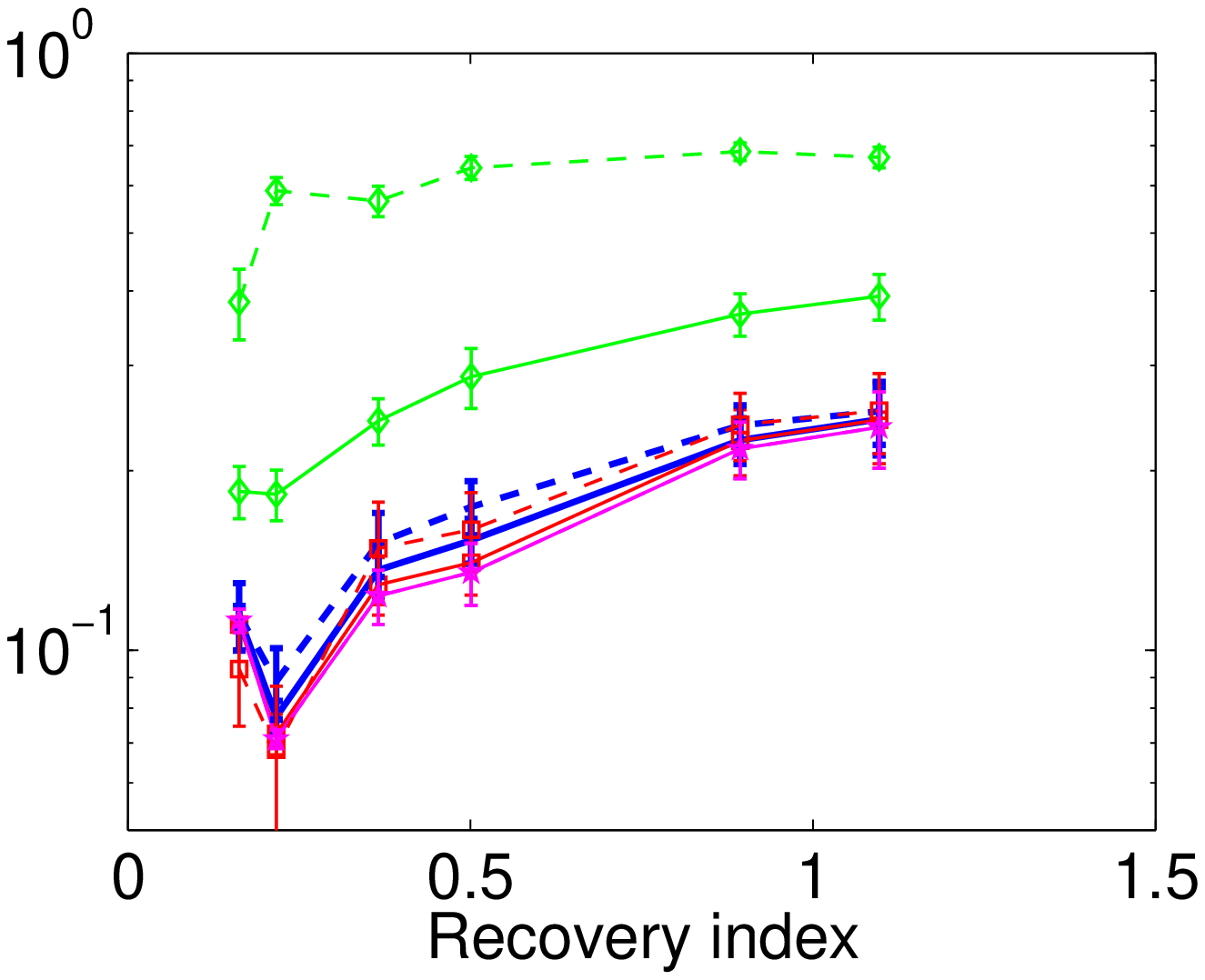}}
\subfloat[1024x2048]{
\includegraphics[width=0.33\textwidth]{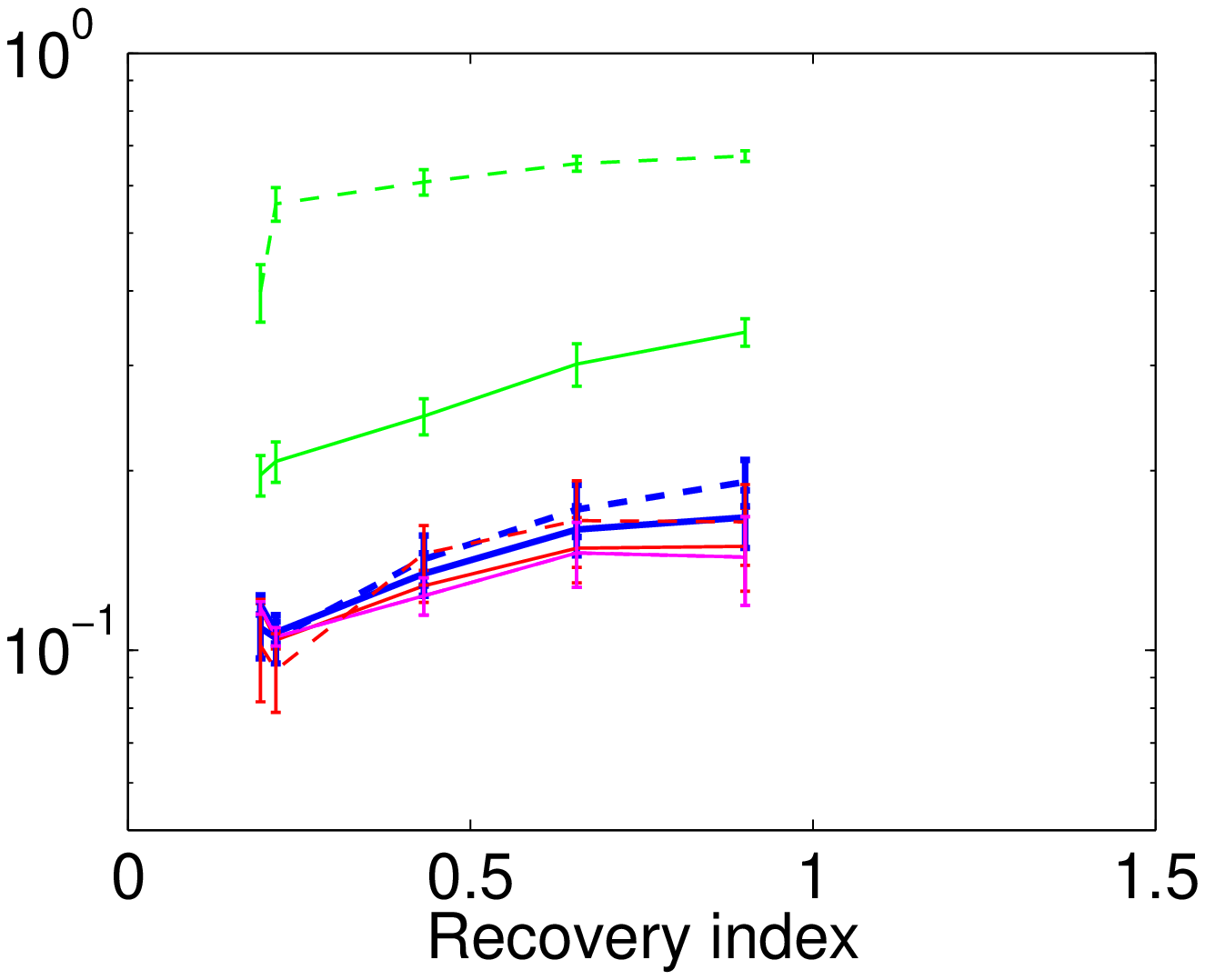}}
\subfloat[2048x4096]{
\includegraphics[width=0.33\textwidth]{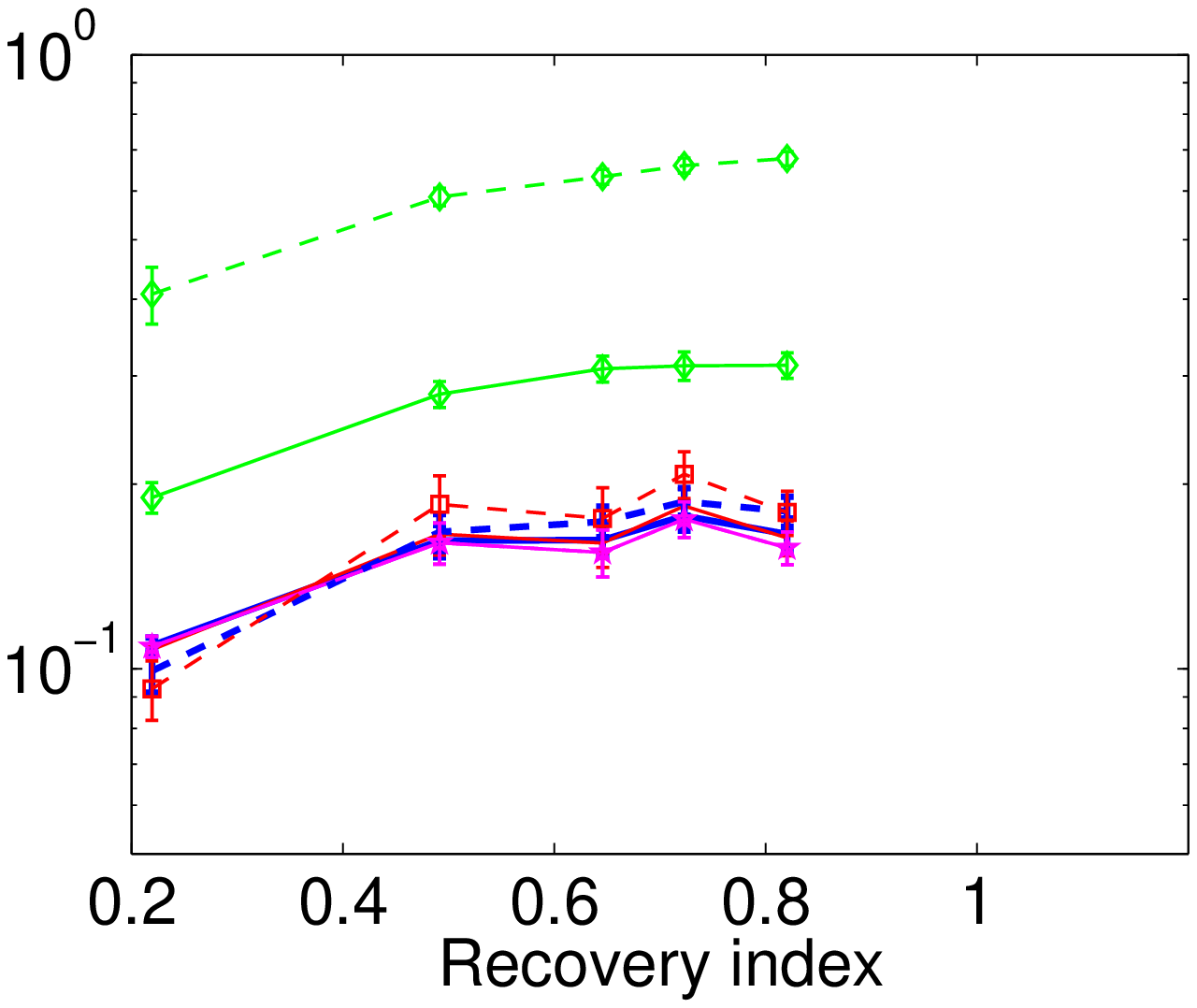}}

\caption{Recovering a DFT from undersampled data. Showing the average performance of the CS
algorithms for various problem dimensions and effective sparseness degrees. The convergence time and the normalized error are depicted
in the upper and lower panels, respectively.}
\label{fig:sm_dft}
\end{figure}

The averaged performance of the various methods over 50 Monte Carlo runs (in which a new set of frequencies $\{w_i\}^{n_f}_{i=1}$
is sampled at the beginning of each run) is depicted in Fig.~\ref{fig:sm_dft}.
The upper panel in this figure shows the CS methods' mean convergence times and their associated standard deviations (error bars)
for different problem dimensions (ranging from 512x1024 through 1024x2048 to 2048x4096) and effective sparseness levels (corresponding
to recovery indices of between $0.1$ to $0.8$). The corresponding recovery errors of the various methods are provided in the
bottom panel of this figure.

By observing both these panels, it can be clearly recognized that the CSP method maintains
the \emph{best tradeoff} between accuracy and computational load as the problem becomes more complex (as indicated by both its
dimensionality and sparseness degree). In virtue of its underlying mechanism, the CSP exhibits robustness with respect to both these factors
as it attains recovery errors that are comparable to those of the BCS and the OMP at a nearly fixed computational cost.
This renders it the fastest reliable method as the problem dimension increases. Although the GP is the fastest scheme in this case, its accuracy is extremely low with recovery errors
of nearly the magnitude of the signal itself. The LS-augmented GP does posses a clear advantage over the unaugmented
one; however, its performance is still unsatisfactory (with recovery errors of almost twice than those of the
other methods).

The above insights are further illustrated in both Tables~\ref{tab:s_dft} and \ref{tab:l_dft} in which the timing
and recovery error values are repeated for the small and large scale scenarios, respectively. Thus, it can be seen from the
first table (small scale) that the CSP attains a recovery error that is similar to those of the BCS and the OMP.
Its convergence time, however, is longer compared to the other methods. The prominent advantage of
the CSP is manifested in the 2nd table (large scale) in which it attains the best performance
in terms of accuracy and convergence time for a relatively large recovery index of $0.8$.
Its robustness to the sparseness level is illustrated in both these tables by the nearly fixed convergence time for the two
extremal values of the recovery index.

\begin{table}[tbh]
\centering
\begin{tabular}{|l|c|c|c|c|c|c|}
\hline\hline
Method & \multicolumn{2}{c|}{$(\hat s/m) \log n=0.2$} & \multicolumn{2}{c|}{$(\hat s/m) \log n=0.8$} & $(\hat s/m) \log n=0.2$ & $(\hat s/m) \log n=0.8$ \\
\hline
OMP  & 0.11  & \textbf{0.11} & 0.21  & \textbf{0.21}  & 0.07 (sec) & 4.19 (sec)\\
BCS  & 0.09  & \textbf{0.11} & 0.23  & \textbf{0.22}  & 1.11 (sec) & 1.86 (sec)\\
GP   & 0.38  & \textbf{0.18} & 0.68  & \textbf{0.36}  & 0.05 (sec) & 0.05 (sec)\\
CSP  & 0.11  & \textbf{0.11} & 0.23  & \textbf{0.22}  & 4.11 (sec) & 4.93 (sec)\\
\hline\hline
\end{tabular}

\caption{Recovering a DFT from undersampled data. The normalized recovery error
(left columns) and convergence time (right columns) of the various methods for the problem dimension 512x1024.
The bold values correspond to the accuracy of the two-staged LS-augmented variants. Averaged over 50 Monte Carlo runs.}
\label{tab:s_dft}
\end{table}

\begin{table}[tbh]
\centering
\begin{tabular}{|l|c|c|c|c|c|c|}
\hline\hline
Method & \multicolumn{2}{c|}{$(\hat s/m) \log n=0.2$} & \multicolumn{2}{c|}{$(\hat s/m) \log n=0.8$} & $(\hat s/m) \log n=0.2$ & $(\hat s/m) \log n=0.8$ \\
\hline
OMP  & 0.10  & \textbf{0.10} & 0.17  & \textbf{0.17}  & 7.20 (sec)  & 226.80 (sec)\\
BCS  & 0.09  & \textbf{0.10} & 0.20  & \textbf{0.18}  & 15.40 (sec) & 50.62 (sec)\\
GP   & 0.40  & \textbf{0.19} & 0.66  & \textbf{0.31}  & 0.47 (sec)  & 0.53 (sec)\\
CSP  & 0.09  & \textbf{0.10} & 0.18  & \textbf{0.17}  & 12.87 (sec) & 13.11 (sec)\\
\hline\hline
\end{tabular}

\caption{Recovering a DFT from undersampled data. The normalized recovery error
(left columns) and convergence time (right columns) of the various methods for the problem dimension 2048x4096.
The bold values correspond to the accuracy of the two-staged LS-augmented variants. Averaged over 50 Monte Carlo runs.}
\label{tab:l_dft}
\end{table}

\begin{figure}[htb]
\centering
\subfloat[Sparse signal]{
\includegraphics[width=0.50\textwidth]{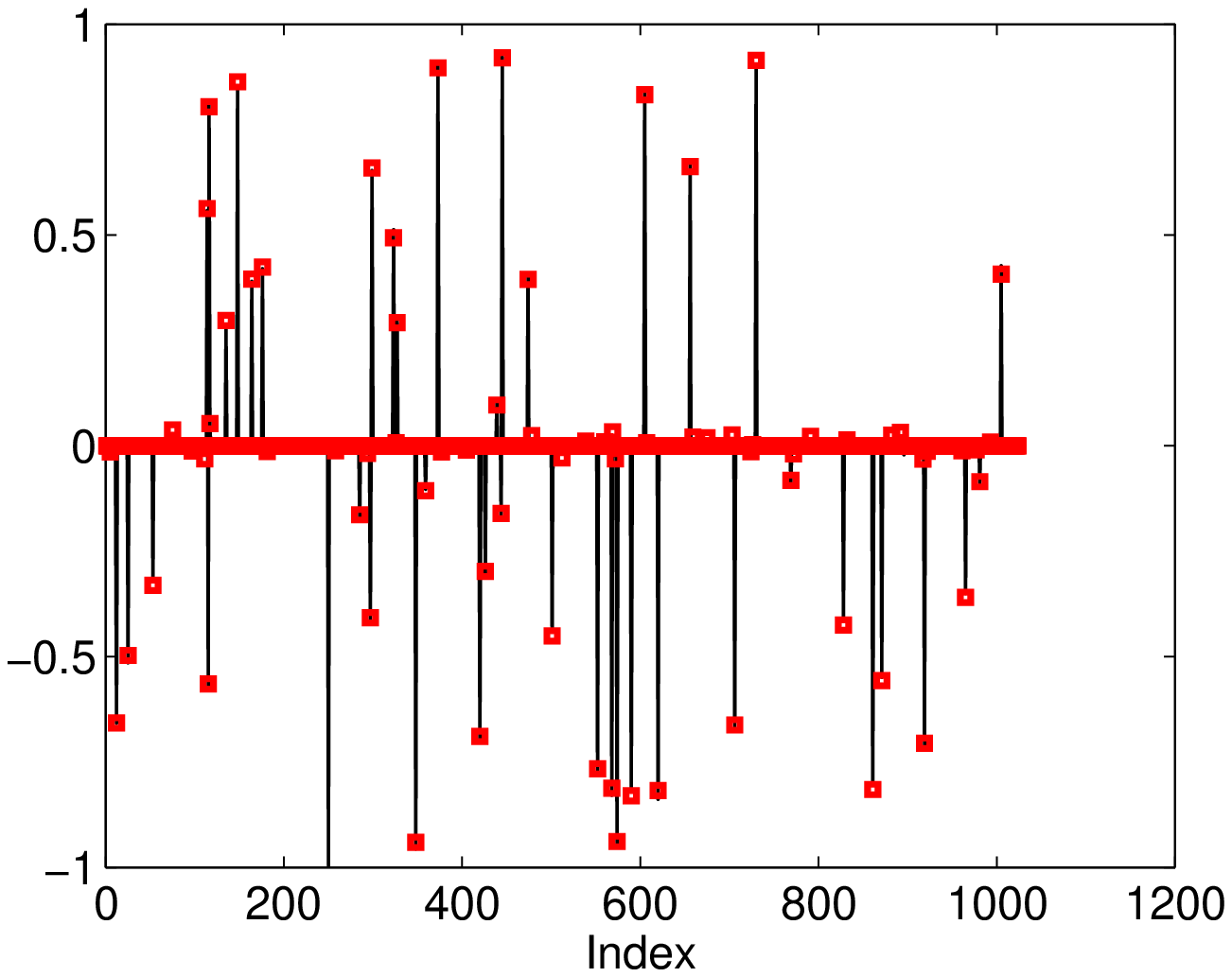}}
\subfloat[DFT (compressible) signal]{
\includegraphics[width=0.50\textwidth]{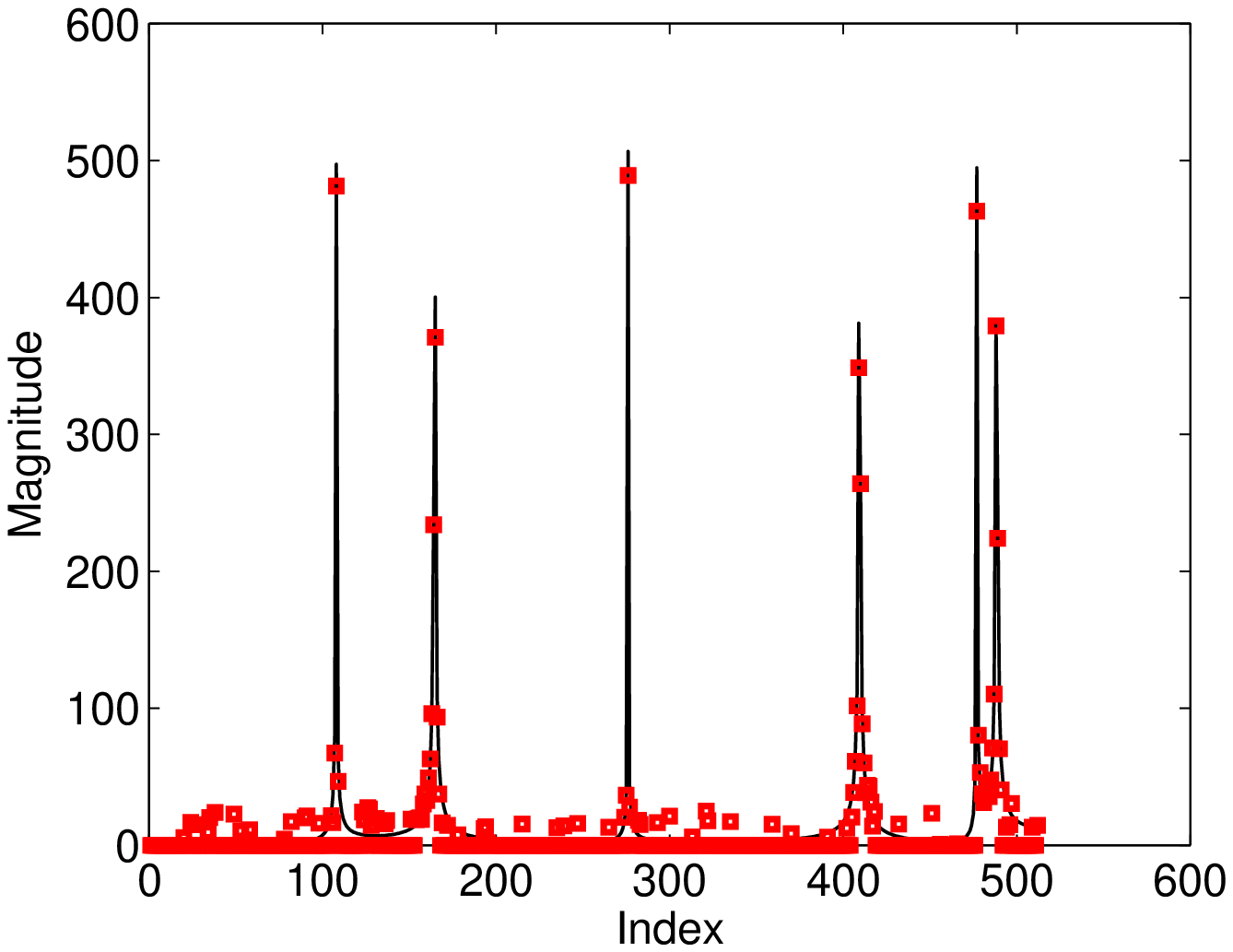}}

\caption{Illustration of the recovery performance of the CSP method for sparse and compressible signals.}
\label{fig:sc}
\end{figure}

Figure~\ref{fig:sc} depicts the performance of the CSP in recovering typical sparse and compressible signals of nearly
the same recovery index. This figure suggests that, in practice, though the recovery indices are nearly the same,
it might be more difficult to reconstruct a compressible representation rather than a sparse one. This
follows from the fact that the compressible estimate on the right sub-figure is less accurate than its companion
on the left.

\subsubsection{Image Recovery from Undersampled Radial Fourier Coefficients}

In the last part of this section, we demonstrate the performance of some of the methods in recovering
an image using undersampled 2D Fourier coefficients that are computed along radial lines. The example
considered here follows along the lines of \cite{dantzig:06}, where the Shepp-Logan phantom head image is used.
In our experiment, however, we use a low-dimensional 128x128 version of this image, which yields a signal of dimension
$128^2=16384$.
We examine the CS methods for two scenarios in which the 2D Fourier coefficients are sampled along either 32 or 64 radial lines
(corresponding to nearly $25\%$ or $50\%$ of the available data).

\begin{figure}[htb]
\centering
\subfloat[Original]{
\includegraphics[width=0.33\textwidth]{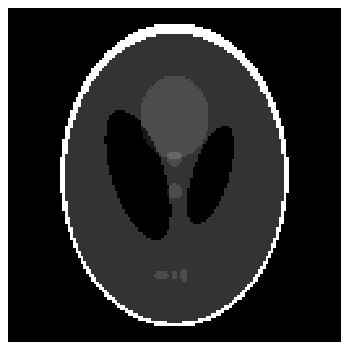}}
\subfloat[Normalized error 0.12]{
\includegraphics[width=0.33\textwidth]{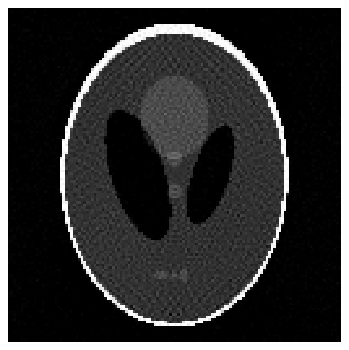}}
\subfloat[Normalized error 0.22]{
\includegraphics[width=0.33\textwidth]{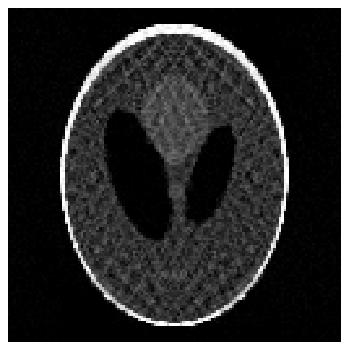}}\\
\subfloat[Original]{
\includegraphics[width=0.33\textwidth]{phantom128}}
\subfloat[Normalized error 0.19]{
\includegraphics[width=0.33\textwidth]{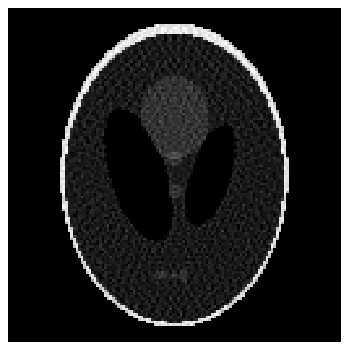}}
\subfloat[Normalized error 0.28]{
\includegraphics[width=0.33\textwidth]{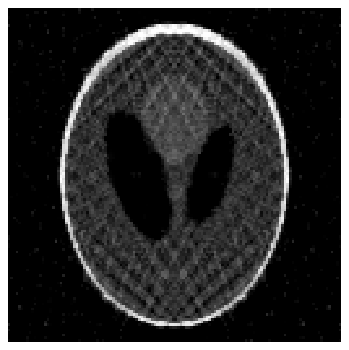}}

\caption{The original and reconstructed 128x128 Shepp-Logan phantom head. Showing
the CSP (upper panel) and the GP (lower panel) methods. The middle and right rows depict the recovery error based on 64 and 32
projections, respectively (equivalent to 50\% and 25\% undersampled data).}
\label{fig:phantom}
\end{figure}

Three methods are applied for recovering the phantom head image: BCS, GP and CSP (unaugmented versions).
However, the performance of only two of them, the GP and the CSP are shown in Fig.~\ref{fig:phantom} as the BCS
exhibited poor performance, essentially yielding recovery errors of nearly the signal magnitude (specifically, around
$0.9$). Figure~\ref{fig:phantom} clearly shows the superiority of the CSP over the GP in both scenarios.



\section{Conclusions}

A novel compressed sensing (CS) scheme was introduced for solving large-scale compressible problems.
The new method utilizes the instantaneous subgradient for projecting the previous iterate on some intermediate point, thereby approaching the underlying convex feasibility set resulting from a group of possibly nonlinear constraints.
The cyclic subgradient projection (CSP) mechanism, which is the heart of the new approach, facilitates
the efficient solution of large-scale CS problems owing to its implementation, which involves vector products only.
An extensive numerical comparison of the CSP-CS algorithm and its variants with some of the state-of-the-art
CS schemes clearly demonstrated its superiority in high-dimensional compressible
settings.

In particular, we may draw the following conclusions. First, the newly-proposed method maintains a nearly fixed computational cost, irrespectively to the problem dimension and sparseness level; moreover, the new method easily copes with realistic scenarios involving high dimensional
compressible signals.

Second, a de-biased CSP, which utilizes an additional least-squares stage, improves the performance for sparse signals, showing superiority compared to other de-biased methods.

Finally, the extensive numerical comparison of all the methods clearly shows
that the CSP maintains the best tradeoff between computational efficiency and accuracy as the problem dimension
and sparseness level increase. As such, we conclude that the CSP is the most efficient method for large-scale compressible problems.

\section*{Acknowledgments}

This research was partially supported by the Asher Space Research Institute. The Authors wish to acknowledge Yair Censor, whose seminal work on optimization inspired the research described herein.


\appendix


The \texttt{MATLAB}\textsuperscript{\textregistered} files implementing our new algorithms are available at\\ \texttt{http://www.technion.ac.il/~pgurfil/csp-cs/}. All the other source files for the various CS algorithms implemented herein can be found at the following locations:

\begin{itemize}
\item Least Angle Regression (LARS) : \\
\texttt{http://www.mathworks.com/matlabcentral/fileexchange/23186-lars-algorithm}
\item Basis Pursuit (BP): \\
\texttt{http://www.acm.caltech.edu/l1magic/}
\item Bayesian CS (BCS): \\
\texttt{http://people.ee.duke.edu/~lihan/cs/}
\item Gradient Projection for Sparse Reconstruction (GPSR): \\
\texttt{http://www.lx.it.pt/~mtf/GPSR/}
\end{itemize}

 \bibliographystyle{ieee}

\begin{thebibliography}{10}

\bibitem{cs:06}
E.~J. Candes, J.~Romberg, and T.~Tao,
\newblock ``Robust uncertainty principles: Exact signal reconstruction from
  highly incomplete frequency information'',
\newblock {\em IEEE Transactions on Information Theory}, vol. 52, pp. 489--509,
  2006.

\bibitem{candes:06}
E.~J. Candes,
\newblock ``Compressive sampling'',
\newblock Madrid, Spain, 2006, European Mathematical Society, Proceedings of
  the International Congress of Mathematicians.

\bibitem{MRI:07}
M.~Lustig, D.~Donoho, and J.~M. Pauly,
\newblock ``Sparse {MRI}: The application of compressed sensing for rapid {MR}
  imaging'',
\newblock {\em Magnetic Resonance in Medicine}, vol. 58, pp. 1182--1195, 2007.

\bibitem{MRI:08}
U.~Gamper, P.~Boesiger, and S.~Kozerke,
\newblock ``Compressed sensing in dynamic {MRI}'',
\newblock {\em Magnetic Resonance in Medicine}, vol. 59, pp. 365--373, 2008.

\bibitem{nonconvex:07}
R.~Chartrand,
\newblock ``Exact reconstruction of sparse signals via nonconvex
  minimization'',
\newblock {\em IEEE Signal Processing Letters}, vol. 14, pp. 707--710, 2007.

\bibitem{tibshirani:96}
R.~Tibshirani,
\newblock ``Regression shrinkage and selection via the {LASSO}'',
\newblock {\em Journal of the Royal Statistical Society. Series B
  (Methodological)}, vol. 58, no. 1, pp. 267--288, 1996.

\bibitem{lars:04}
B.~Efron, T.~Hastie, I.~Johnstone, and R.~Tibshirani,
\newblock ``Least angle regression'',
\newblock {\em Annals of Statistics}, vol. 32, no. 2, pp. 407 -- 499, 2004.

\bibitem{dantzig:06}
E.~Candes and T.~Tao,
\newblock ``The {Dantzig} selector: statistical estimation when p is much
  larger than n'',
\newblock {\em Annals of Statistics}, vol. 35, pp. 2313--2351, 2007.

\bibitem{bp:98}
S.~S. Chen, D.~L. Donoho, and M.~A. Saunders,
\newblock ``Atomic decomposition by basis pursuit'',
\newblock {\em SIAM Journal of Scientific Computing}, vol. 20, no. 1, pp. 33 --
  61, 1998.

\bibitem{rvm:01}
M.~E. Tipping,
\newblock ``Sparse {Bayesian} learning and the relevance vector machine'',
\newblock {\em Journal of Machine Learning Research}, vol. 1, pp. 211 -- 244,
  2001.

\bibitem{mc1}
R.~E. McCulloch and E.~I. George,
\newblock ``Approaches for {Bayesian} variable selection'',
\newblock {\em Statistica Sinica}, vol. 7, pp. 339 -- 374, 1997.

\bibitem{mc2}
J.~Geweke,
\newblock {\em Bayesian Statistics 5}, chapter Variable selection and model
  comparison in regression,
\newblock Oxford University Press, 1996.

\bibitem{mc3}
B.~A. Olshausen and K.~Millman,
\newblock ``Learning sparse codes with a mixture-of-{Gaussians} prior'',
\newblock {\em Advances in Neural Information Processing Systems (NIPS)}, pp.
  841 -- 847, 2000.

\bibitem{mc4}
S.~J. Godsil and P.~j. Wolfe,
\newblock ``Bayesian modelling of time-frequency coefficients for audio signal
  enhancement'',
\newblock {\em Advances in Neural Information Processing Systems (NIPS)}, 2003.

\bibitem{bcs}
S.~Ji, Y.~Xue, and L.~Carin,
\newblock ``Bayesian compressive sensing'',
\newblock {\em IEEE Transactions on Signal Processing}, vol. 56, pp. 2346 --
  2356, June 2008.

\bibitem{cskf}
A.~Carmi, P.~Gurfil, and D.~Kanevsky,
\newblock ``Methods for sparse signal recovery using kalman filtering with
  embedded pseudo-measurement norms and quasi-norms'',
\newblock {\em IEEE Transactions on Signal Processing. Accepted}.

\bibitem{mp}
S.~Mallat and Z.~Zhang,
\newblock ``Matching pursuits with time-frequency dictionaries'',
\newblock {\em IEEE Transactions on Signal Processing}, vol. 4, pp. 3397 --
  3415, 1993.

\bibitem{omp}
Y.~C. Pati, R.~Rezifar, and P.~S. Krishnaprasad,
\newblock ``Orthogonal matching pursuit: recursive function approximation with
  applications to wavelet decomposition'',
\newblock 27th Asilomar Conf. on Signals, Systems and Comput., 1993.

\bibitem{ols}
S.~Chen, S.~A. Billings, and W.~Luo,
\newblock ``Orthogonal least squares methods and their application to
  non-linear system identification'',
\newblock {\em International Journal of Control}, vol. 50, pp. 1873 -- 1896,
  1989.

\bibitem{gp}
M.~A.~T. Figueiredo, R.~D. Nowak, and S.~J. Wright,
\newblock ``Gradient projection for sparse reconstruction: Application to
  compressed sensing and other inverse problems'',
\newblock {\em IEEE Journal of Selected Topics in Signal Processing}, vol. 1,
  pp. 586 -- 597, December 2007.

\bibitem{c93}
P.~L. Combettes,
\newblock ``The foundations of set-theoretic estimation'',
\newblock {\em Proceedings of the IEEE}, vol. 81, pp. 182--208, 1993.

\bibitem{c96}
P.~L. Combettes,
\newblock ``The convex feasibility problem in image recovery'',
\newblock {\em Advances in Imaging and Electron Physics}, vol. 95, pp.
  155--270, 1996.

\bibitem{H80}
G.~T. Herman,
\newblock {\em Image reconstruction from projections: The fundamentals of
  computerized tomography},
\newblock Academic Press, New York, NY, USA, 1980.

\bibitem{cap88}
Y.~Censor, M.~D. Altschuler, and W.~D. Powlis,
\newblock ``On the use of {C}immino's simultaneous projections method for
  computing a solution of the inverse problem in radiation therapy treatment
  planning'',
\newblock {\em Inverse Problems}, vol. 4, pp. 607--623, 1988.

\bibitem{msl99}
L.~D. Marks, W.~Sinkler, and E.~Landree,
\newblock ``A feasible set approach to the crystallographic phase problem'',
\newblock {\em Acta Crystallographica}, vol. A55, pp. 601--612, 1999.

\bibitem{butnariu}
D.~Butnariu, Y.~Censor, P.~Gurfil, and E.~Hadar,
\newblock ``{On the behavior of subgradient projections methods for convex
  feasibility problems in Euclidean spaces}'',
\newblock {\em SIAM Journal on Optimization}, vol. 19, no. 2, pp. 786--807,
  2008.

\bibitem{yamada}
I.~Yamada,
\newblock ``{Hybrid steepest descent method for variational inequality problem
  over the fixed point set of certain quasi-nonexpansive mappings}'',
\newblock {\em Numerical Functional Analysis and Optimization}, vol. 25, pp.
  619--655, 2004.

\bibitem{crombez03}
G.~Crombez,
\newblock ``{Non-monotoneous parallel iteration for solving convex feasibility
  problems}'',
\newblock {\em Kybernetika}, vol. 39, pp. 547--560, 2003.

\bibitem{crombez06}
G.~Crombez,
\newblock ``{A sequential iteration algorithm with non-monotoneous behaviour in
  the method of projections onto convex sets}'',
\newblock {\em Czechoslovak Mathematical Journal}, vol. 56, pp. 491--506, 2006.

\bibitem{CZ97}
Y.~Censor and S.~A. Zenios,
\newblock {\em {Parallel optimization: Theory, algorithms, and applications}},
\newblock Oxford University Press, New York, NY, USA, 1997.

\bibitem{lc82}
Y.~Censor and A.~Lent,
\newblock ``{Cyclic subgradient projections}'',
\newblock {\em Mathematical Programming}, vol. 24, pp. 233--235, 1982.

\bibitem{moledo}
A.~N. Iusem and L.~Moledo,
\newblock ``A finitely convergent method of simultaneous subgradient
  projections for the convex feasibility problem'',
\newblock {\em Computational and Applied Mathematics}, vol. 5, pp. 169--184,
  1986.

\bibitem{dossantos}
L.~T. Dos~Santos,
\newblock ``A parallel subgradient method for the convex feasibility problem'',
\newblock {\em Journal of Computational and Applied Mathematics}, vol. 18, pp.
  307--320, 1987.

\end{thebibliography}

\end{document}